\documentclass[structabstract]{aa}  
\usepackage{graphicx}
\usepackage{textcomp}
\usepackage{epsfig}
\usepackage{longtable,lscape}
\usepackage{txfonts}
\usepackage{mathrsfs}
\usepackage[normalem]{ulem}
\usepackage[round]{natbib}
\usepackage[]{hyperref}
\begin{document}
   \title{The AMIGA sample of isolated galaxies}

   \subtitle{XII. Revision of the isolation degree for AMIGA galaxies using the SDSS\thanks{Tables 2 and 4 are only available in electronic form at the CDS via anonymous ftp to cdsarc.u-strasbg.fr (130.79.128.5) or via \texttt{http://cdsweb.u-strasbg.fr/cgi-bin/qcat?J/A+A/}. Initial tables are available at the AMIGA webpage \texttt{http://amiga.iaa.es} .}}

   \author{M.\,Argudo-Fern\'andez
          \inst{1,2}
          \and
          S.\,Verley\inst{2}
          \and
          G.\,Bergond\inst{3}
          \and
          J.\,Sulentic\inst{1}
          \and
          J.\,Sabater\inst{4}
          \and
          M.\,Fern\'andez\,Lorenzo\inst{1}
          \and
          S.\,Leon\inst{5}
          \and
          D.\,Espada\inst{5,6,7}
          \and
          L.\,Verdes-Montenegro\inst{1}
          \and
          J.\,D.\,Santander-Vela\inst{1}
          \and
          J.\,E.\,Ruiz\inst{1}
          \and
          S.\,S\'anchez-Exp\'osito\inst{1}
          }

   \institute{Instituto de Astrof\'isica de Andaluc\'ia (CSIC) Apdo. 3004, 18080 Granada, Spain
         \and
             Departamento de F\'isica Te\'orica y del Cosmos, Universidad de Granada, 18071 Granada, Spain
         \and
             Centro Astron\'omico Hispano Alem\'an, Calar Alto, (CSIC-MPG), C/ Jes\'us Durb\'an Rem\'on 2-2, E-04004 Almer\'ia, Spain    
         \and
             Institute for Astronomy, University of Edinburgh, Edinburgh EH9 3HJ, UK
         \and
             Joint ALMA Observatory (ALMA/ESO), Alonso de C\'ordova 3107, Vitacura, Santiago 763-0355, Chile 
         \and
             National Astronomical Observatory of Japan (NAOJ), 2-21-1 Osawa, Mitaka, Tokyo 181-8588, Japan
         \and      
             Department of Astronomical Science, The Graduate University for Advanced Studies (SOKENDAI), 2-21-1 Osawa, Mitaka, Tokyo 181-8588, Japan           
             }

   \date{Received February 20, 2013; accepted September 13, 2013}

 
\abstract
{To understand the evolution of galaxies, it is necessary to have a reference sample where the effect of the environment is minimized and quantified. 
In the framework of the AMIGA project (\textbf{A}nalysis of the interstellar \textbf{Me}dium of \textbf{I}solated \textbf{GA}laxies), we present a revision of the environment for galaxies in the \textbf{C}atalogue of \textbf{I}solated \textbf{G}alaxies (CIG, Karachentseva 1973) using the ninth data release of the Sloan Digital Sky Survey (SDSS-DR9).}
{The aims of this study are to refine the photometric-based AMIGA sample of isolated galaxies and to provide an improvement of the quantification of the isolation degree with respect to previous works, using both photometry and spectroscopy.}
{We developed an automatic method to search for neighbours within a projected area of 1\,Mpc radius centred on each primary galaxy to revise the CIG isolation criteria introduced by Karachentseva (1973). The local number density at the fifth nearest neighbour and the tidal strength affecting the CIG galaxy were estimated to quantify the isolation degree.}
{Of the 636 CIG galaxies considered in the photometric study, 426 galaxies fulfil the CIG isolation criteria within 1\,Mpc, taking into account projected neighbours. Of the 411 CIG galaxies considered in the spectroscopic study, 347 galaxies fulfil the CIG isolation criteria when a criterion about redshift difference is added. The available redshifts allow us to reject background neighbours and thus improve the photometric assessment. On average, galaxies in the AMIGA sample show lower values in the local number density and the tidal strength parameters than galaxies in denser environments such as pairs, triplets, compact groups, and clusters.}
{For the first time, the environment and the isolation degree of AMIGA galaxies are quantified using digital data. The use of the SDSS database permits one to identify fainter and smaller-size satellites than in previous AMIGA works. The AMIGA sample is improved by this study, because we reduced the sample of isolated galaxies used in previous AMIGA works by about 20\%. The availability of the spectroscopic data allows us to check the validity of the CIG isolation criteria, which is not fully efficient. About 50\% of the neighbours considered as potential companions in the photometric study are in fact background objects. We also find that about 92\% of the neighbour galaxies that show recession velocities similar to the corresponding CIG galaxy are not considered by the CIG isolation criteria as potential companions, which may have a considerable influence on the evolution of the central CIG galaxy.}


   \maketitle
%

\section{Introduction}  \label{Sec:intro}

During the past forty years it has become clear that galaxy properties and evolution can be driven as much by environment as by initial conditions, even if the details of environmental influence are not yet well quantified. Some observed properties show a strong dependence on the environment, for instance, on the optical and ultraviolet luminosity and atomic gas mass function \citep{2009ARA&A..47..159B}, on the infrared luminosity function \citep{1991ApJ...374..407X} or the associated stellar mass function \citep{2001ApJ...557..117B}, on the morphology-mass relation \citep{2011MNRAS.tmpL.354C}, and on the galaxy colours \citep{2004ApJ...601L..29H,2006MNRAS.373..469B}.

Isolated galaxies are located in environments of such low density that they have not been appreciably affected by their closest neighbours during a past crossing time $t_{\rm{cc}} = $\,3\,Gyr \citep{2005A&A...436..443V}. The observed physical properties of these systems are expected to be mainly determined by initial formation conditions and secular evolutionary processes. A representative sample of isolated galaxies is therefore needed to test models of galaxy formation and evolution. It may also serve as a reference sample in studies of galaxies in pairs, triplets, groups, and clusters. This will aid our understanding of the effects of the environment on fundamental galaxy properties.

Statistical studies of isolated galaxies require a large and morphologically diverse sample. Few good samples of isolated galaxies exist, one of the largest being the Catalogue of Isolated Galaxies \citep[CIG;][]{1973AISAO...8....3K}. The original visual systematic search for isolated galaxies using the First Palomar Observatory Sky Survey (POSS-I) employs a visual projected isolation criterion. Since the redshift distances of only a few galaxies were known at that time, isolation in the third dimension could not be directly estimated. Instead, any galaxy with nearby similar-size neighbours was rejected. The resulting CIG includes 1050 galaxies (plus CIG 781, a globular cluster mistakenly included in the original list). 

Despite the importance of analysing pure 'nature' samples, not many additional studies of isolated galaxies were carried out in the following three decades \citep{1977ApJ...216..694H,1980AJ.....85.1010A,1981Afz....17...53A,1982ApJ...253..526B,2004A&A...420..873V}. This led many scientists to assume that no real isolated galaxy population exists. It is natural therefore that the AMIGA (\textbf{A}nalysis of the interstellar \textbf{M}edium of \textbf{I}solated \textbf{GA}laxies\footnote{\texttt{http://amiga.iaa.es}}) project \citep{2005A&A...436..443V} is based upon a re-evaluation of the CIG. It is a first step in trying to identify and better understand isolated galaxies in the local Universe. \citet{2008MNRAS.390..881D} used a subsample of 100 typical CIG galaxies (Sb, Sbc, and Sc) and found that most of them have a bulge-to-total luminosity ratio $B/T~<$~0.1. If $B/T$ is a measure of environmental dynamical processing \citep{2010ApJ...709L..53M}, galaxies in the CIG sample appear to be very little affected 
by it. The late-type population that dominates the AMIGA sample \citep{2006A&A...449..937S} may indicate that they have been alone for most of their lives.

\citet{2007A&A...472..121V,2007A&A...470..505V} calculated two isolation parameters (the local number density and the tidal strength) for 950 galaxies in the CIG sample using an automated search for neighbours on the first and second digitised POSS (DPOSS-I and II) based on photographic plates. They provided an exhaustive list of $\sim$\,54,000 possible satellites that were used to identify several CIG galaxies failing the CIG isolation criteria. 

The first data release of the Sloan Digital Sky Survey \citep[SDSS-DR1;][]{2003AJ....126.2081A} has rekindled the interest on isolated galaxy studies in the past decade \citep{2005AJ....129.2062A}. In this context, new and available photometric data from the SDSS motivated us to perform a fully digital revision of the isolation degree for the CIG galaxies. The SDSS-III \citep{2011AJ....142...72E} maps one third of the sky using CCD detectors. The SDSS also provides spectra that allow us to estimate galaxy distances, enabling an improved revision of the degree of isolation for 411 CIG galaxies with a fairly complete spectroscopic coverage in the last Data Release \citep[DR9;][]{2012ApJS..203...21A} catalogue. Some other recent catalogues of isolated galaxies have been compiled, introducing isolation criteria using the spectroscopic information from earlier SDSS data releases. \citet{2009MNRAS.394.1409E} applied three-dimensional Voronoi tessellation to volume-limited galaxy samples, using spectroscopic data 
from SDSS-DR5, and identified 2394 isolated galaxies. \citet{2009AN....330.1004V} refined the sample by selecting galaxies with the highest level of isolation, the QIsol sample, composed of 600 galaxies. These two samples suffer from the incompleteness in the SDSS spectroscopic sample, limited at $m_{\rm{r,Petrosian}} < 17.77$\,mag. To compensate for the SDSS spectroscopic incompleteness, other authors used statistical techniques \citep{2012MNRAS.424.1454E}, photometric redshifts also provided by the SDSS \citep{2011MNRAS.417..370G}, or selected different volume-limited samples \citep{2011ApJ...738..102T}. A revision for possible photometric companions is needed out of the volume-limited samples considered \citep{2010AJ....139.2525H}.
    
Here we perform a photometric and spectroscopic census and quantify the environment of the CIG galaxies covered by the SDSS-DR9. In Sect.~\ref{Sec:amiga}, we present the CIG, as well as the revisions and improvements on isolation performed within the AMIGA project. In Sect.~\ref{Sec:data}, we describe in detail the data and methodology used to revise the isolation of the CIG galaxies in the SDSS, including a description of our automated pipeline used to produce a catalogue of their potential neighbours. The method to quantify the isolation degree, as well as the selection of the comparison samples used, are explained in Sect.~\ref{Sec:isolparam}. Results from the study using the photometric and spectroscopic catalogues of the SDSS are presented in Sect.~\ref{Sec:result}. A revision of the CIG is presented in Sect.~\ref{Sec:diss} to determine how many galaxies remain isolated based on the recent SDSS-DR9 data from both the photometric and the spectroscopic catalogues. Neighbours considered in each study are 
then used for the estimation of the isolation degree. We present our conclusions in Sect.~\ref{Sec:con}.


\section{AMIGA project} \label{Sec:amiga}

The AMIGA project adopts the Catalogue of Isolated Galaxies \citep[CIG;][]{1973AISAO...8....3K} as a starting point and proceeds to extract a refined sample of the historically most significant sample of isolated galaxies in the local Universe.

All CIG galaxies are found in the Catalogue of Galaxies and Clusters of Galaxies \citep[CGCG;][]{1968cgcg.bookR....Z} with apparent photographic magnitudes $m_{\rm{pg}} < 15.7$\,mag. These very isolated systems represent $\sim$\,3\,\% of the CGCG.
   
The CIG isolation criteria (Eqs.~\ref{Eq:kara2} and \ref{Eq:kara1}) consider a primary galaxy of angular diameter $D_P$ as isolated if there is no neighbour $i$ with an angular diameter $D_{i}$ between 0.25 and $4\times D_P$ lying within a projected distance 20 times the diameter of the neighbour:

\begin{equation} \label{Eq:kara2}
\frac{1}{4} \,D_{P} \leq D_{i} \leq 4 \,D_{P}\quad;
\end{equation} 

\begin{equation} \label{Eq:kara1}
R_{iP} \geq 20 \,D_{i}\quad.
\end{equation} 

The AMIGA project refines the pioneering CIG in several ways, including a revision of all galaxy positions \citep{2003yCat..34110391L}, an optical study, including sample redefinition, magnitude correction, and full-sample analysis of the optical luminosity function \citep{2005A&A...436..443V}, a morphological revision and type-specific optical luminosity function analysis \citep{2006A&A...449..937S}, a study on H{$\alpha$} morphology \citep{2007A&A...474...43V}, and a re-evaluation of the degree of isolation of the CIG \citep{2007A&A...472..121V,2007A&A...470..505V}. The original CIG contains 1051 items, but one of the compiled objects is a globular cluster \citep[CIG 781;][]{2005A&A...436..443V} so the size of the sample considered in the rest of this paper is N = 1050.
     
The AMIGA project also started several multiwavelength studies for galaxies in the CIG: characterisation of the $B$-band luminosity function \citep{2006A&A...449..937S}; Fourier photometric decomposition, optical asymmetry, and photometric clumpiness and concentration \citep{2008MNRAS.390..881D,2009MNRAS.397.1756D}; characterisation of the FIR luminosity function \citep{2007A&A...462..507L}, radio-continuum \citep{2008A&A...485..475L}, molecular gas \citep{2012A&A...538C...1L}, and atomic gas \citep{2011A&A...532A.117E}; characterisation of nuclear activity \citep{2008A&A...486...73S,2012A&A...545A..15S}; optical colours \citep{2012A&A...540A..47F}; and optical study of the stellar mass-size relation \citep{2013MNRAS.434..325F}.

\subsection{Previous revision of the CIG environment}   \label{Sec:sample}

One of the AMIGA improvements of the CIG involves the revision and quantification of the CIG isolation criteria. \citet{2007A&A...470..505V} used DPOSS-I and DPOSS-II images for this revision. The digitised images from photographic plates enabled them to revise the environment description for all 950 CIG galaxies with radial velocities higher than 1500\,km\,s$^{-1}$ within a minimum physical radius of 0.5\,Mpc. All neighbour candidates brighter than $m_{B}~=~17.5$\,mag were identified in each field with a fair degree of confidence, using the LMORPHO software \citep{1995PASP..107..770O, 1996ApJ...472L..13O, 2002ApJ...568..539O}. A catalogue of approximately 54,000 neighbours was created, and redshifts are available for only $\sim$30\% of this sample.

\citet{2007A&A...472..121V} used two complementary parameters to quantify the isolation degree of the CIG galaxies, the local number density of neighbour galaxies $\eta_{k}$, and the tidal strength $Q$ affecting the central galaxy by its neighbourhood. The local number density $\eta_{k}$ is defined as follows:
\begin{equation} \label{Eq:etak}
\eta_{k} \propto {\rm log}\left(\frac{k - 1}{V(r_{k})}\right)\quad,
\end{equation} 
where $V(r_{k}) = \frac{4}{3}\,\pi\,r_{k}^{3}$ and $r_{k}$ is the projected distance to the $k^{\rm{th}}$ nearest neighbour, with $k$ equal to 5 or lower if there are not enough neighbours in the field. And the tidal strength exerted by one companion is defined as
\begin{equation} \label{Eq:Qip}
Q_{iP} \equiv \frac{F_{\rm{tidal}}}{F_{\rm{bind}}} 
\propto {\frac{M_{i}}{M_{P}}}\left(\frac{D_{P}}{R_{iP}}\right)^{3}\quad,
\end{equation} 
where $M_{i}$ and $M_{P}$ are the mass of the neighbour and the primary galaxy, respectively, $D_{P}$ is the apparent diameter of the primary galaxy, and $R_{iP}$ the projected distance between the neighbour and primary galaxy. Using the apparent diameter as an approximation for galaxy mass,
\begin{equation} \label{Eq:Q2007}
Q_{iP} \equiv \frac{F_{\rm{tidal}}}{F_{\rm{bind}}} 
\propto \left(\frac{\sqrt{D_{P}D_{i}}}{R_{iP}}\right)^{3}\quad.
\end{equation}
This approximation is based on the dependence of galaxy mass $M$ on size: $M \varpropto D^{\gamma}$, with $\gamma = 1.5$ \citep{1984AJ.....89..966D,2004ApJ...604..521T}. The final tidal parameter considered is a dimensionless estimation of the gravitational interaction strength, calculated from the logarithm of the sum of the tidal strengths created by all the neighbours in the field, $Q=$\,log$( \sum Q_{iP} )$. 
 
In this paper we calculate modified improved versions of these two parameters for the CIG using photometry and spectroscopy from the SDSS (see Sect.~\ref{Sec:isolparam}).
\section{Data and methodology}         \label{Sec:data}

The SDSS-DR9\footnote{\texttt{http://www.sdss3.org/}} \citep{2011AJ....142...72E,2012ApJS..203...21A} provides images and spectra covering $14,555$ square degrees mostly of the northern sky. The SDSS database provides homogeneous and moderately deep photometry in five pass-bands. The 95\% completeness limits for the images are $(u,g,r,i,z) = (22.0, 22.2, 22.2, 21.3, 20.5)$, respectively. The images are mostly taken under good/average seeing conditions (the median is about 1$\farcs$4 in $r$-band) on moonless nights. The photometric catalogue of detected objects was used to identify the targets for spectroscopy: a) the main galaxy sample \citep{2002AJ....124.1810S}, with a target magnitude limit of $m_{r,\rm{Petrosian}}~<~17.77$\,mag corrected for Galactic dust extinction, and b) the Baryon Oscillation Spectroscopic Survey \citep[BOSS;][]{2013AJ....145...10D}, which uses a new spectrograph \citep{2012arXiv1208.2233S} to obtain spectra of galaxies with $0.15 < z < 0.8$ and quasars with $2.15 < z < 3.5$, which 
is useful to reject background objects in our study. The data are processed using automatic pipelines \citep{2011AJ....142...31B}. 

\subsection{CIG galaxies in the SDSS}  \label{Sec:CIGselect}

\begin{figure*}
\centering
\includegraphics[width=.9\textwidth]{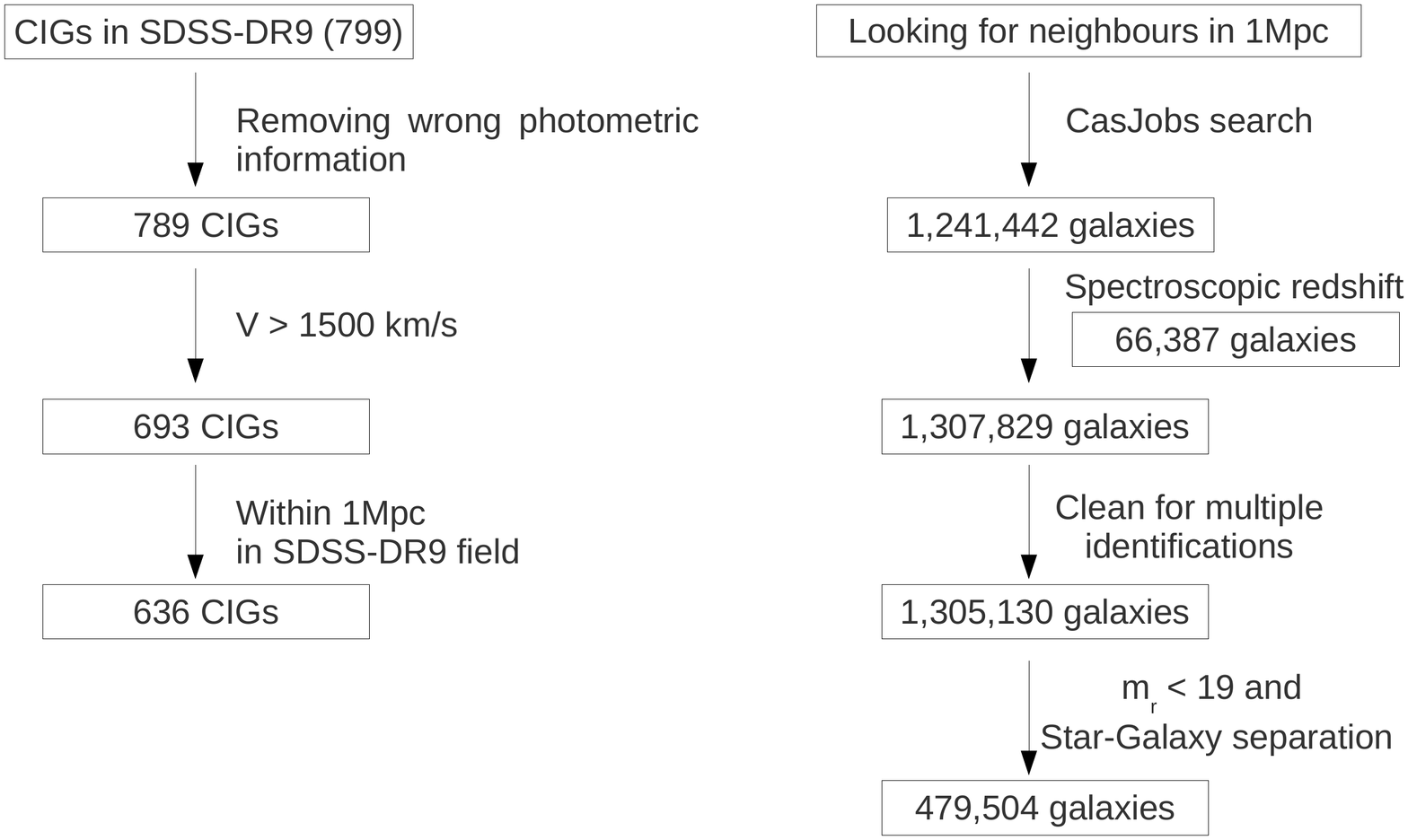}
\caption[Diagram of the methodology]{Diagram of the methodology. The scheme used to select primary galaxies is shown in the left column, and the selection for the neighbours is shown in the right column.}
\label{Fig:pipeline}
\end{figure*}

The scheme of the pipeline we followed is presented in Fig.~\ref{Fig:pipeline}. We found $N = $~799 CIG galaxies included in the SDSS photometric catalogue, of which ten were removed because the photometric data were unreliable (due to a nearby bright star or because the galaxy is too close to an edge of the field): CIG 13, 95, 388, 402, 573, 713, 736, 781, 802, and 810. We used recession velocities from the AMIGA database\footnote{\texttt{http://amiga.iaa.es/p/139-amiga-public-data.htm}} \citep{2012A&A...540A..47F} for CIG galaxies. We chose a projected physical radius of 1\,Mpc ($H_{0}$=75\,km\,s$^{-1}$\,Mpc$^{-1}$) to evaluate the isolation degree\footnote{Note that the search radius is larger than in \citet{2007A&A...470..505V}, who used a minimum physical radius of 0.5\,Mpc due to technical limitations.}. If we were to assume a typical field velocity dispersion of the order of 190\,km\,s$^{-1}$ \citep{2000ApJ...530..625T} it would require about $t_{\rm{cc}}\sim\,5.2$\,Gyr for a companion to cross this 
distance, guaranteeing that the galaxy has been isolated most of its lifetime. We focused our study on CIG galaxies with recession velocities $\varv \geq 1500$\,km\,s$^{-1}$, which additionally reduced our sample to $N = $~693, to avoid an overwhelming search for potential neighbours (the angular size on the sky for 1\,Mpc at a distance of 1500\,km\,s$^{-1}$ is approximately 2{\textdegree}.9).

The CIG isolation criteria (Eqs.~\ref{Eq:kara2} and \ref{Eq:kara1}) require one to examine the isolation in a field as large as 80 times the diameter of each CIG galaxy. For a typical CIG galaxy, with diameter $D_{P} = 30$\,kpc, this translates into a projected distance of $R_{iP} = 2.4$\,Mpc. Even selecting a reasonable and constant search radius, the variable radius resulting from the CIG isolation criteria usually represents a very large field, and 1\,Mpc often corresponds to corresponds to a part of it. Therefore, we cannot verify the isolation for the entire field used by \citet{1973AISAO...8....3K}, but we are able to determine if neighbour galaxies close to the primary CIG galaxy are violating the CIG isolation criteria.

Model magnitudes in $r$-band (the deepest images) were used in our study. We used $r_{90}$, the Petrosian radius containing 90\,\% of the total flux of the galaxy in the $r$-band\footnote{\texttt{http://www.sdss3.org/dr9/algorithms/magnitudes.php}}, as explained in Sect.~\ref{Sec:diameter}.

Our final sample is composed of $N = $~636 CIG galaxies, and 1\,Mpc radius fields are completely covered in the photometric SDSS-DR9 catalogue.

\subsection{Catalogues of neighbours}  \label{Sec:neighselect}
   
We used the CasJobs\footnote{\texttt{http://skyservice.pha.jhu.edu/CasJobs/}} tool to search for neighbour galaxies within a 1\,Mpc radius around each of the 636 CIG galaxies\footnote{To allow the reproducibility of this work, initial tables are available at http://amiga.iaa.es, CDS, and SDSS-DR9 websites and requests on demand.} (see right column in Fig.~\ref{Fig:pipeline}). Neighbour galaxies were selected with the following criteria to extract a sample as clean as possible: 1) galaxies with $11.0 \leq m_{r} \leq 21.0$ without flags on size measures, 2) removal of suspicious detections, checking that the object has pixels detected in the first pass and has a valid radial profile, and 3) flagged as a non-saturated source.

A first sample of 1,241,442 candidate neighbour galaxies was compiled using these conditions. Without imposing the condition for non-saturated objects, we find a contamination of nearly 50\% by saturated stars with magnitudes brighter than $m_{r} \sim 17$\,mag. Galaxies with a very bright nucleus can also be flagged as saturated sources in the SDSS, which makes it necessary to complete our sample by adding saturated galaxies from the spectroscopic catalogue (66,387 galaxies). Our final sample contains 1,307,829 neighbour galaxies selected by an automated method from the SDSS.

We found a contamination of multiple object identifications for nearby and extended galaxies. A clean sample of 1,305,130 galaxies was obtained selecting the brightest (typically also the largest) object in these cases. 

\begin{figure}
\centering
\includegraphics[width=0.49\textwidth]{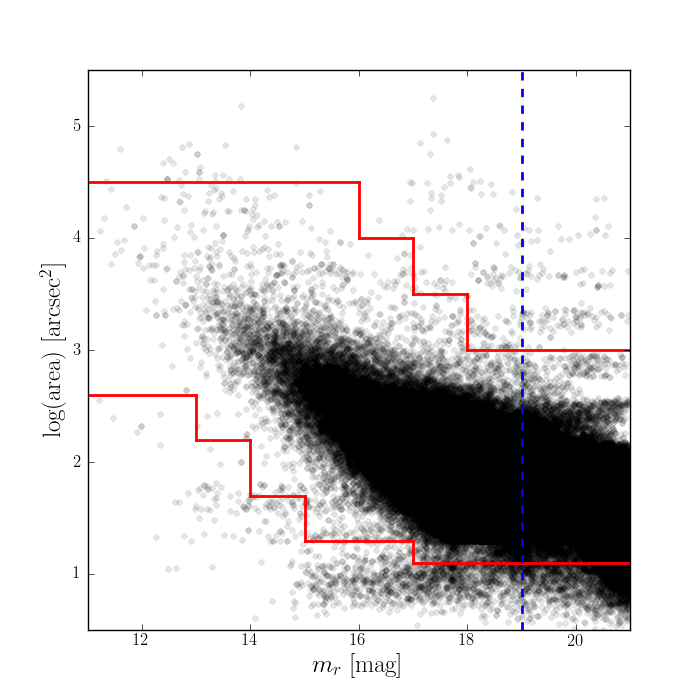}
\caption[Star-galaxy separation]{Star-galaxy separation. Considering the object area from the Petrosian radius $r_{90}$, we carried out an empirical inspection, performing a selection in apparent magnitude and size. Resolved objects within the red contour are very likely galaxies. The dashed blue line corresponds to the selected cut at $m_{r} = 19$\,mag.}
\label{Fig:SGsep}
\end{figure}

We also improved the star-galaxy separation provided by the SDSS from an empirical selection of objects using a size/magnitude diagram (see Fig.~\ref{Fig:SGsep}). Objects situated in the horizontal bottom part are mostly stars misclassified as galaxies. Bright objects in the upper part of the diagram are saturated stars, fainter objects in the upper part (and below log[Area]~=~1.1) are spurious detections. Objects fainter than $m_{r} = 19$\,mag were removed, since the star-galaxy separation becomes difficult and inaccurate. Our final photometric sample of neighbour galaxies is composed of a total of 479,504 neighbour galaxies around 636 CIG galaxies.

\subsection{Estimation of apparent diameters from the SDSS} \label{Sec:diameter}

The CIG isolation criteria defined by \citet{1973AISAO...8....3K} are based on apparent diameters of galaxies, which makes these measurements critical in our study. SDSS gives different radius measurements (for the five photometric bands): 1) de Vaucouleurs and exponential radii, which depend directly on the galaxy intensity profile, and 2) Petrosian radii $r_{\rm{Petrosian}}$, using a modification of the \citet{1976ApJ...209L...1P} system. Petrosian values measure galaxy fluxes within a circular aperture and define the radius using the shape of the azimuthally averaged light profile. Petrosian radii containing 90\% ($r_{90}$) and 50\% ($r_{50}$) of the total flux are provided by the SDSS. We adopted Petrosian values for this study because they do not depend on model fits \citep{2010MNRAS.404.2087B}. However, after a visual inspection of SDSS three-colour images for CIG galaxies, Petrosian diameters do not recover the total galaxy major axis, and are generally smaller than the projected major axis of a 
galaxy at the 25 mag/arcsec$^{2}$ isophotal level ($D_{25}$) originally used by Karachentseva (values measured in the $B$-band), even though a new approach for background subtraction was applied in the last two data releases of the SDSS \citep{2011AJ....142...31B}. 
   
To transform Petrosian sizes into more accurate optical measurements equivalent to the original $D_{25}$ used by Karachentseva, we compared Petrosian diameters from the SDSS ($D_{\rm{SDSS}} = 2r_{90}$) with apparent optical diameters given in ancillary databases. We performed a linear regression analysis for CIG galaxies using measures from 1) HyperLeda\footnote{\texttt{http://leda.univ-lyon1.fr/}}; 2) $isoA_{\rm{r}}$ (isophotal major axis at the 25 mag/arcsec$^2$ isophote in $r$-band) from SDSS-DR7 \citep{2009ApJS..182..543A} because we visually verified that in general it covers the total galaxy; 3) the major axis from the NASA/IPAC Extragalactic Database (NED\footnote{\texttt{http://ned.ipac.caltech.edu/}}); and 4) the Kron radii, running SExtractor \citep{1996A&AS..117..393B} on $i$-band images of CIG galaxies from the SDSS-DR9, which, after visually examining on the fits images, is the measurement that better recovers the total size of the CIG galaxy.

But CIG galaxies are not representative of the size of the galaxies that we are interested in: neighbour galaxies are typically smaller than CIG galaxies. To avoid bias, we focused the size correction on neighbour galaxies. We performed a cross-match with the catalogue of neighbours compiled by \citet{2007A&A...470..505V}, based on $D_{25}$ and SExtractor diameters. Since this correlation is made from digitised photographic measurements, other correlations were calculated: based on $D_{25}$ from HyperLeda for galaxies in the neighbour sample brighter than $m_{r,\rm{model}}=16$\,mag and with $r_{90} > 5$\,arcsec; and based on the SExtractor Kron radii from SDSS-DR9 images in the $i$-band for four CIG fields at different recession velocities. 

Correction factors for SDSS diameters as a function of the above measurements are shown in Table~\ref{tab:Dsdss}, including the corresponding number of galaxies used. In the rest of this study, we use a corrected apparent diameter $D_{\rm{SDSS,\,corr}} = 1.43\,D_{\rm{SDSS}}$ \citep[see also][]{2013MNRAS.430..638S} both for neighbours and for CIG galaxies. This factor was obtained as the median of the values in Table~\ref{tab:Dsdss} to approximate the original $D_{25}$ used by \citet{1973AISAO...8....3K}.

\begin{table}
  \caption[Estimation of apparent diameters from the SDSS]{Estimation of apparent diameters from the SDSS using different 
   comparison samples. 
   Col. 1: Galaxy samples considered for the estimation.
   Col. 2: Apparent diameter measure used for the estimation. 
   Col. 3: Apparent diameter source. 
   Col. 4: Number of objects used. 
   Col. 5: Correction factor for each estimation $D_{25}\simeq \rm{factor}\times D_{\rm{SDSS}}$.}
   \label{tab:Dsdss}
\centering
\begin{tabular}{lllcc}
\hline \hline
  Objects & Apparent & Database & \# of & Factor \\
   & diameter & & matches & \\
\hline 
  CIGs & $D_{25}$ & HyperLeda & 636 & 1.58 \\
  CIGs & $isoA_{\rm{r}}$ & SDSS-DR7 & 560 & 1.41 \\
  CIGs & Major axis & NED & 567 & 1.83 \\
  CIGs & Kron radius & SDSS-DR9 & 719 & 2.17 \\
\hline
  Neighbours & $D_{25}$ & Verley+07c & 28,209 & 1.41 \\
  Neighbours & $D_{25}$ & HyperLeda & 13,972 & 1.23 \\
  Neighbours & Kron radius & SDSS-DR9 & 27,719 & 1.43 \\
\hline   
\end{tabular}
\end{table}

\section{Quantification of the isolation} \label{Sec:isolparam}

\subsection{Isolation parameters} \label{Sec:defQeta}

We used modified isolation parameters from Sect.~\ref{Sec:amiga} to quantify the isolation degree throughout. 
   
An estimate of $\eta_{k}$ was calculated taking into account the distance of the $k^{\rm{th}}$ nearest neighbour to the CIG galaxy. We calculated this parameter according to Eq.~\ref{Eq:etak}, choosing $k$ equal to 5 or lower when there were not enough neighbours in the field. The farther the $k^{\rm{th}}$ nearest neighbour, the smaller the local number density $\eta_{k}$.

We calculated a second independent parameter involving a cumulative measure of the tidal strength produced by neighbour galaxies. To improve the quantification of the isolation degree, we adopted a modified version of the $Q$ parameter (Eq.~\ref{Eq:Qip}), where apparent magnitudes from the SDSS-DR9 were used to estimate galaxy masses. This methodology therefore minimises the effect of the correction factor used for estimating apparent diameters. Assuming that the stellar mass is proportional to the $r$-band flux, that is a linear mass-luminosity relation \citep{2003ApJS..149..289B,2006ApJ...652..270B}, we considered $Flux_{r}\propto \mathscr{M}ass$ at a fixed distance, with $m_{r}$~=~$-$2.5\,log$(Flux_{r})$. 
   
Then, for one companion from Eq.~\ref{Eq:Qip}:
\begin{equation} \label{Eq:Q2012}
{\rm log}Q_{iP} \propto 0.4\,(m_{r}^{P}-m_{r}^{i}) + 3\,{\rm log}\left(\frac{D_{P}}{R_{iP}}\right)\quad,
\end{equation} 
where $m_{r}^{P}$ and $m_{r}^{i}$ are the apparent magnitudes in $r$-band of the primary CIG galaxy and the $i^{\rm{th}}$ neighbour, respectively. The total tidal strength created by all the neighbours is then defined as
\begin{equation} \label{Eq:Q2012tot}
Q = {\rm log}\left(\sum_{i}Q_{iP}\right)\quad.
\end{equation}

The higher the value of $Q$, the more affected from external influence the galaxy, and viceversa.

Given that the CIG is assembled with the requirement that no similar size neighbours are found close to the CIG galaxy, companion galaxies are expected to be faint (mostly dwarf companions). Therefore, no companion galaxies with similar brightness are expected close to the CIG galaxy.
   
\subsection{Comparison with denser environments} \label{Sec:denser}

We selected other samples of galaxies from denser environments to make a comparison with the isolation degree of galaxies in the CIG sample: 1) isolated pairs of galaxies \citep[KPG; ][]{1972SoSAO...7....1K}, 2) galaxy triplets \citep[KTG; ][]{1979AISAO..11....3K}, 3) galaxies in compact groups \citep[HCG; ][]{1982ApJ...255..382H}, and 4) galaxies in Abell clusters \citep[ACO; ][]{1958ApJS....3..211A,1989ApJS...70....1A}.

The KPG sample allowed us to separate effects of galaxy environment density from effects of one-on-one interactions, while the KTG, HCG, and ACO galaxy samples show the effects of increasingly richer environments. The KPG, KTG, and HCG catalogues were visually compiled using visual isolation criteria as well, accordingly, they complement the CIG sample nicely.

The KTG, HCG, and ACO samples were adopted from \citet{2007A&A...472..121V} because they were selected to sample a volume of space roughly equivalent to the one covered by the CIG, and to avoid possible biases. For consistency, we also followed the same selection criterion as for CIG galaxies, keeping galaxies with recession velocities $\varv \geq 1500$\,km\,s$^{-1}$.

The final comparison sample is composed of 360 KPGs out of 603 pairs listed by \citet{1972SoSAO...7....1K}, 30 KTGs out of 84 triplets listed by \citet{1979AISAO..11....3K}, 24 HCGs out of 100 compact groups compiled by \citet{1982ApJ...255..382H}, and 12 ACOs out of more than 2,700 galaxy clusters listed by \citet{1958ApJS....3..211A} and \citet{1989ApJS...70....1A}. 
   
\section{Results} \label{Sec:result}

We performed a photometric and, for the first time, spectroscopic revision of the CIG isolation criteria around each CIG galaxy within a projected radius of 1\,Mpc. The available redshift information allowed us to identify possible physical companions down to the SDSS spectroscopic completeness. Neighbour galaxies were used to estimate the isolation degree for the CIG galaxies. 

\subsection{Photometric study} \label{Sec:resultphoto}   

\subsubsection{Photometric revision of the CIG isolation criteria} \label{Sec:resultphotokara}

We applied both of the CIG isolation criteria to the neighbours around each CIG galaxy within a projected field radius of 1\,Mpc. First of all, we identified neighbour galaxies with an apparent diameter in the range defined by Eq.~\ref{Eq:kara2}. They represent the galaxies considered as potential perturbers by \citet{1973AISAO...8....3K}. The second step was to determine which of these objects were projected at a distance lower than the one defined in Eq.~\ref{Eq:kara1}. When a galaxy was found to have no neighbour violating Eqs.~\ref{Eq:kara2} and \ref{Eq:kara1}, that galaxy was considered isolated according to the CIG isolation criteria.

In Sect.~\ref{Sec:neighselect} we have compiled an automatic selected sample of 479,504 neighbours; of these, 2,109 are potential companions according to the CIG isolation criteria (Eqs.~\ref{Eq:kara2} and \ref{Eq:kara1}) within 1\,Mpc. After an additional visual inspection we found that 89 candidates fail our aim of obtaining a sample of neighbour galaxies without contamination by saturated stars.

The revision of the CIG isolation criteria was performed for 636 CIG galaxies using 479,415 neighbour galaxies within 1\,Mpc radius around each CIG galaxy. Of these, 121,872 neighbour galaxies violate Eq.~\ref{Eq:kara2} within 1\,Mpc, and a small number, 3,433 neighbour galaxies, violate Eq.~\ref{Eq:kara1} within 1\,Mpc. The total number of potential companions violating Eqs.~\ref{Eq:kara2} and \ref{Eq:kara1} within 1\,Mpc is very small, 2,020 galaxies. 

We found 86 CIG galaxies without a neighbour, 117 CIG galaxies with one possible companion, and 433 CIG galaxies with more than one possible companion after applying the CIG isolation criteria within 1\,Mpc. There are 13 CIG galaxies with more than ten possible companions. CIG 589 has the largest number of companions (14 possible companions). 

The search radius of 1\,Mpc for each of the 636 CIG galaxies considered in the photometric study (Sect.~\ref{Sec:resultphoto}), covers the area defined by \citet{1973AISAO...8....3K} for 59 fields only; of these, four CIG galaxies are isolated (CIG 50, 299, 651, and 1032) according to the CIG isolation criteria. 

The results of this revision are listed in columns 2, 3, and 4 of Table~\ref{tab:resultphoto}.

\begin{table}
  \caption[Revision of the isolation degree using photometric data]{Revision of the isolation degree using photometric data.} 
   \label{tab:resultphoto}
\centering
\begin{tabular}{cccccc}
\hline \hline
(1) & (2) & (3) & (4) &  (5) & (6) \\
 CIG & $r_{1\rm{Mpc}}$ & $\frac{r_{1\rm{Mpc}}}{r_{80D_{P}}}$ & isol & $Q_{\rm{Kar, p}}$ & $\eta_{k,\rm{p}}$ \\
\hline
    1     &     35.32     &      0.31     &     0     &    -3.35     &     2.03    \\
    2     &     36.92     &      0.49     &     1     &    -3.19     &     1.70    \\
    4     &    111.58     &      0.65     &     0     &    -2.78     &     1.92    \\
    5     &     32.78     &      0.62     &     0     &    -1.60     &     3.02    \\
    6     &     56.94     &      0.71     &     0     &    -2.84     &     2.34    \\
    7     &     20.22     &      0.26     &     1     &    -3.37     &     2.11    \\
    8     &     40.65     &      0.49     &     0     &    -1.83     &     2.65    \\
\ldots & \ldots & \ldots & \ldots & \ldots  & \ldots \\
\hline   
\end{tabular}
\tablefoot{The full table is available in electronic form at http://amiga.iaa.es and in CDS. The columns correspond to (1) the galaxy identification according to the CIG, (2) the projected angular radius, in arcmin, equal to the adopted distance at 1\,Mpc, and (3) the ratio between projected angular radius at 1\,Mpc (in arcmin) over the original field radius used in the CIG isolation criteria. When the ratio is greater than or equal to 1, the fixed physical radius of 1\,Mpc covers the entire original area. (4) Result of the CIG isolation criteria: ''1'' if the galaxy passes, ''0'' if it fails. (5) $Q_{\rm{Kar, p}}$, tidal strength estimate of similar-size neighbours. (6) $\eta_{k,\rm{p}}$, local number density of similar-size neighbours.}
\end{table}

\subsubsection{Photometric isolation parameters}  \label{Sec:resultphotoparam}
   
\begin{figure*}
\centering
\includegraphics[width=\textwidth]{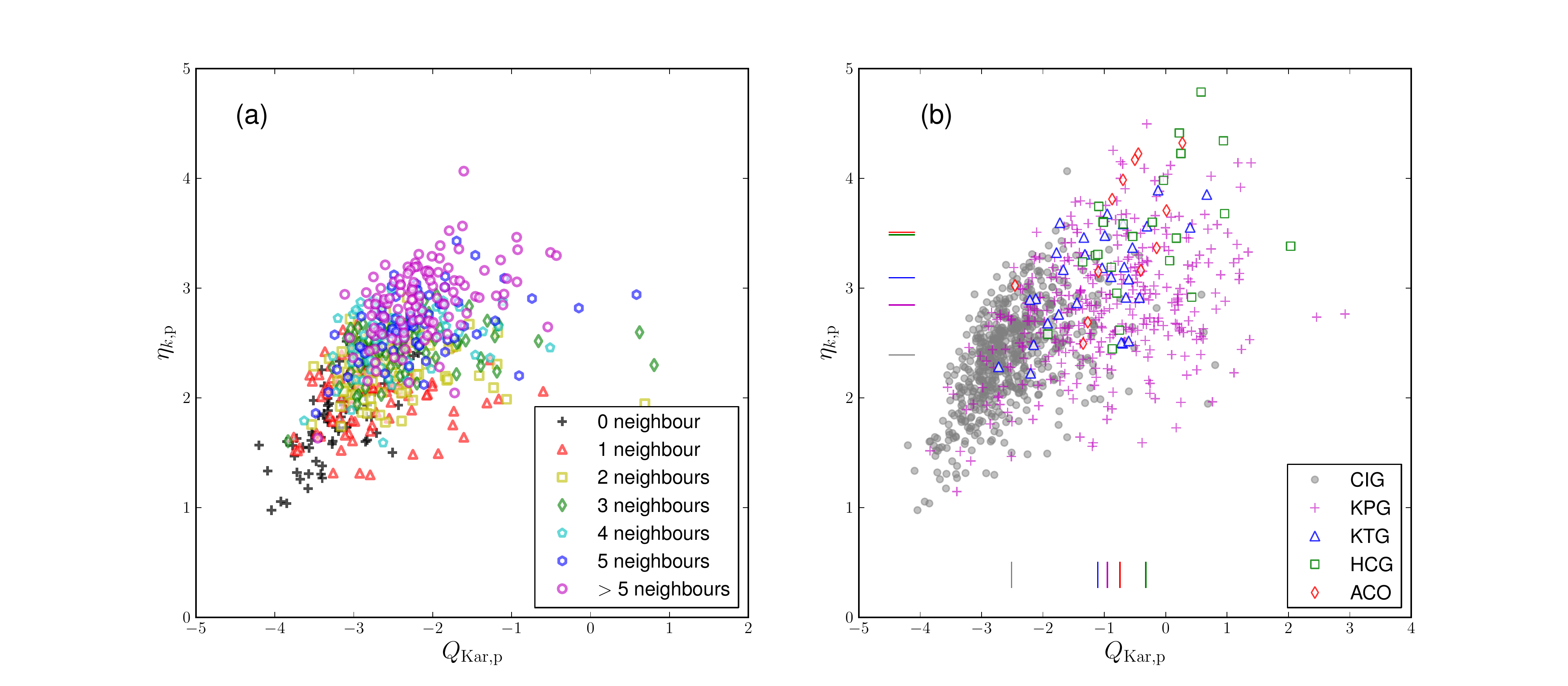}
\caption[Photometric isolation parameters]{Photometric isolation parameters. {\it (a):} Calculated isolation parameters (local number density $\eta_{k,\rm{p}}$ and tidal strength $Q_{\rm{Kar,p}}$) for similar-size neighbour galaxies using the photometric data. Symbols and colours in the legend correspond to the number of neighbours that violate the CIG isolation criteria. {\it (b):} Comparison between isolation parameters (local number density $\eta_{k,\rm{p}}$ and tidal strength $Q_{\rm{Kar,p}}$) for the CIG and the comparison samples using photometric data. Pairs (KPG) are depicted by violet pluses, triplets (KTG) by blue triangles, compact groups (HCG) by green rectangles, and Abell clusters (ACO) by red diamonds. The mean values of each sample are indicated following the same colour code.}
\label{Fig:photoparam}
\end{figure*}

The isolation parameters local number density $\eta_{k,\rm{p}}$ (Eq.~\ref{Eq:etak}) and tidal strength $Q_{\rm{Kar,p}}$ (Eq.~\ref{Eq:Q2012tot}) were calculated using the photometric data. Only galaxies within a factor 4 in apparent diameter with respect to the CIG galaxy were considered (Eq.~\ref{Eq:kara2}) to minimise the contamination of background/foreground galaxies, following \citet{1973AISAO...8....3K}. We calculated ${R_{iP}}$ using projected angular distances on the sky. For the local number density, the projected distance to the $k^{\rm{th}}$ nearest neighbour, $r_{k}$, was calculated as the angular separation in arcmin normalised by the apparent diameter of the central CIG galaxy. The values of the isolation parameters are listed in Table~\ref{tab:resultphoto} and are plotted in Fig.~\ref{Fig:photoparam}a.
   
The tidal strength $Q_{\rm{Kar,p}}$ and the local number density $\eta_{k,\rm{p}}$, were also calculated for the comparison samples KPG, KTG, HCG, and ACO (see Fig.~\ref{Fig:photoparam}b). Means and standard deviations are shown in Table~\ref{tab:compdenser}. As expected, the trend of the mean values, from isolated to denser environments, shows that isolation parameters are sensitive enough to the effects of the environment.

\begin{table}
\caption[Means and standard deviations of the isolation parameters for the CIG and for the comparison samples]{Means and standard deviations of the isolation parameters for the CIG and for the comparison samples.} \label{tab:compdenser}
\centering
\begin{tabular}{lrrrrr}
\hline \hline
   & CIG & KPG & KTG & HCG & ACO \\
\hline 
N                         &      636  &      360  &       30  &        24  &        12 \\
mean($Q_{\rm{Kar, p}}$)  &   $-$2.51  &  $-$0.95  &  $-$1.11  &   $-$0.32  &   $-$0.75 \\
std($Q_{\rm{Kar, p}}$)   &      0.68  &     1.11  &     0.79  &      0.89  &      0.70 \\
mean($\eta_{k,\rm{p}}$)  &      2.39  &     2.85  &     3.09  &      3.49  &      3.51 \\
std($\eta_{k,\rm{p}}$)   &      0.45  &     0.56  &     0.46  &      0.57  &      0.59 \\
\hline   
\end{tabular}
\end{table}

\subsection{Spectroscopic study} \label{Sec:resultspec}

\subsubsection{Spectroscopic revision of the CIG isolation criteria} \label{Sec:resultspeckara}

Owing to the relatively new availability of large spectroscopic surveys, most of the environment has long been estimated only with photometric analysis for any type of samples (isolated, pairs, triplets, groups, and clusters). Only during the past decade and despite the inhomogeneity and incompleteness of the spectroscopic surveys at very low and high redshifts, some spectroscopic studies were performed \citep{2009MNRAS.394.1409E,2009AN....330.1004V}.

In this section we performed a spectroscopic revision and an improvement of the CIG isolation criteria. Eqs.~\ref{Eq:kara2} and \ref{Eq:kara1} within 1\,Mpc were applied to identify the CIG galaxies that appear to be physically isolated using the spectroscopic  sample of the SDSS.
   
For this study, we selected fields with a redshift completeness greater than 80\% with respect to the photometric sample at $m_{r} \leq 17.7$\,mag (the percentage of extended neighbours down to $m_r<17.7$\,mag lying within a 1\,Mpc projected separation from the CIG galaxy that have a measured redshift), which is, approximately, the redshift completeness limit of the SDSS spectroscopic main galaxy sample \citep{2002AJ....124.1810S}. Four hundred and eleven CIG fields fulfil this requirement, surrounded by 70,169 cleaned neighbour galaxies with spectroscopic information.
   
To evaluate the physical association of the projected neighbours, we introduce a third condition based on the velocity difference between neighbour galaxies with respect to each CIG galaxy $|\Delta\,\varv|= \varv_{i} - \varv_{P}$. 

\begin{figure*}
\centering
\includegraphics[width=\textwidth]{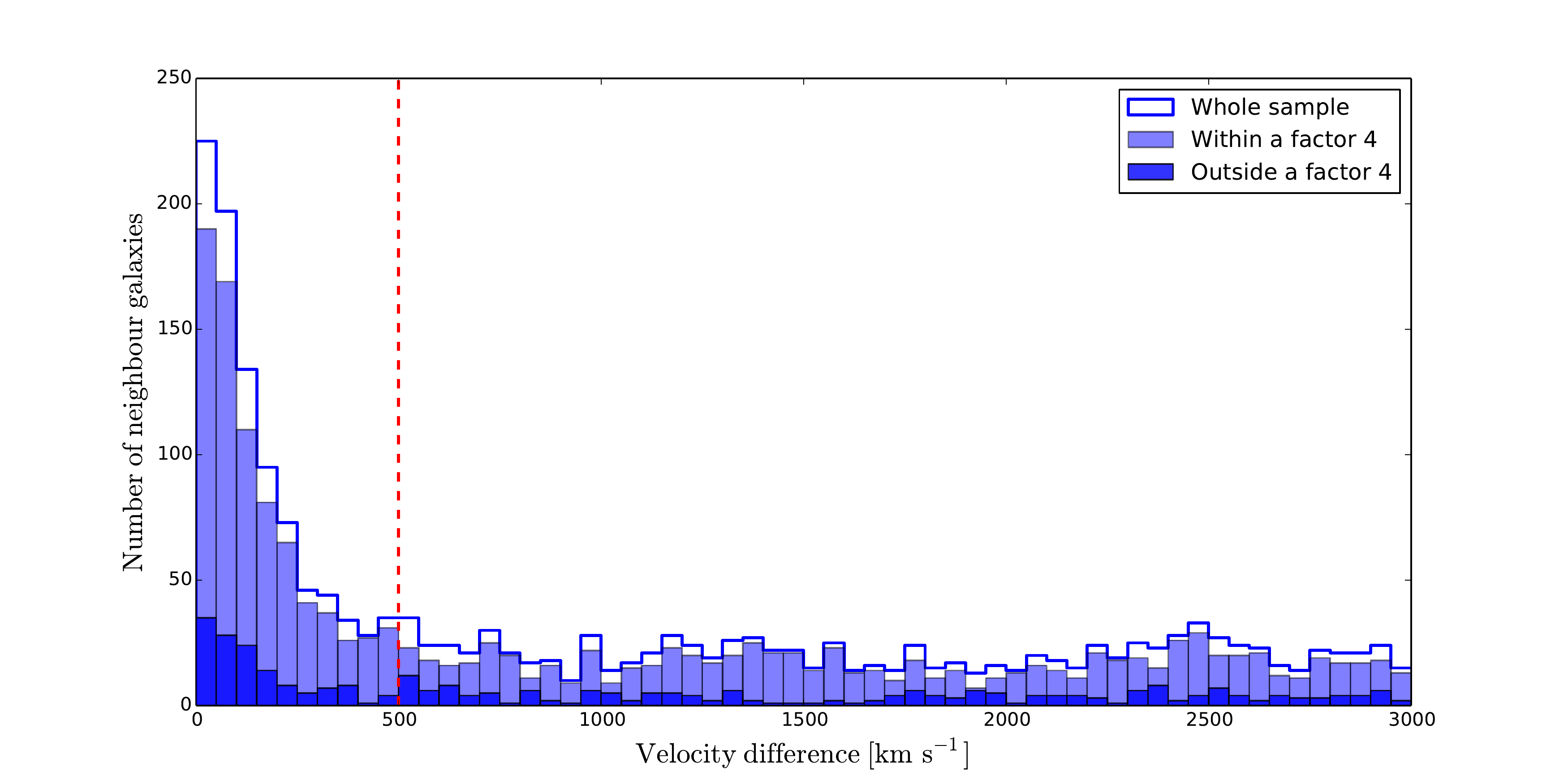}
\caption[Velocity difference distributions]{Comparison of the velocity difference distributions $|\Delta\,\varv|$ for neighbour galaxies with respect to the central CIG galaxy (411 fields): for the neighbour galaxies violating Eq.~\ref{Eq:kara2} (i.e., a factor 4 in apparent diameter with respect to their associated CIG galaxy); for the remaining neighbours (outside the factor 4 in apparent diameter); and for the whole sample of neighbours (sum of the previous two samples). The vertical line corresponds to the selected value of reference at $|\Delta\,\varv| = 500$\,km\,s$^{-1}$.}
\label{Hist:diffvel}
\end{figure*}

Surprisingly, the velocity difference distribution shows a peak close to $|\Delta\,\varv|~=~0$\,km\,s$^{-1}$ (see Fig.~\ref{Hist:diffvel}). From the figure, we are able to separate a flat continuum distribution of foreground/background neighbours, considered as the fraction of galaxies that are probably not linked to the central galaxy, from physically linked satellites. More than one third of the neighbours within $|\Delta\,\varv|~\leq~3,000$\,km\,s$^{-1}$ have a velocity difference $|\Delta\,\varv|~\leq~250$\,km\,s$^{-1}$ (36\%, see Fig.~\ref{Hist:diffvel}). To recover all of these probable physical companions, we considered from the figure and adopting a conservative enough velocity difference selection, that a CIG galaxy fulfils the CIG isolation criteria if it has no neighbour violating Eqs.~\ref{Eq:kara2} and \ref{Eq:kara1} within 1\,Mpc and $|\Delta\,\varv|~\leq~500$\,km\,s$^{-1}$ \citep[see also][]{2009MNRAS.394.1409E,2010Ap.....53..462K,2011AstBu..66..389K}.

The results of the revision of the CIG isolation criteria, using spectroscopic data, are listed in column 2 of Table~\ref{Tab:resultspec}.

The number of galaxies that appear as isolated increases when the third condition is introduced, because they have spectroscopic neighbours with discordant redshifts and violating Eqs.~\ref{Eq:kara2} and \ref{Eq:kara1} within 1\,Mpc. We found that 347 CIG galaxies appear to be isolated according to the CIG isolation criteria and have no companion within 1\,Mpc and $|\Delta\,\varv|~\leq~500$\,km\,s$^{-1}$. The search radius of 1\,Mpc covers the area defined by \citet{1973AISAO...8....3K} for 35 fields only. Of these, 32 CIG galaxies pass, while 3 fail the CIG isolation criteria within 1\,Mpc and $|\Delta\,\varv|~\leq~500$\,km\,s$^{-1}$ (CIG 264, 480, and 637).

A total of 30,222 neighbour galaxies of the 70,169 galaxies with available redshift within the spectroscopic magnitude limit violate Eq.~\ref{Eq:kara2} within 1\,Mpc, and a very small number, 643 neighbours, also violate Eq.~\ref{Eq:kara1} within 1\,Mpc, 75 of them also fulfil the third condition, that is $|\Delta\,\varv|~\leq~500$\,km\,s$^{-1}$ within 1\,Mpc.
       
\begin{table*}
  \caption[Revision of the isolation degree using spectroscopic data]{Revision of the isolation degree using spectroscopic data.} 
   \label{Tab:resultspec}
\centering
\begin{tabular}{ccccccccccccc}
\hline \hline
(1) & (2) & (3) & (4) & (5) & (6) & (7) & (8) & (9) & (10) & (11) & (12) & (13)\\ 
 CIG  &  isol  &  $k_{500}$  &  $Q_{500}$  &  $f_{Q_{500}}$  &  $\eta_{k,500}$ &  $f_{\eta_{k,500}}$  &  $z_{\rm{comp}}$\,[\%]  &  $k_{500,\rm{ul}}$  &  $Q_{500,\rm{ul}}$  &  $f_{Q_{500,\rm{ul}}}$  &  $\eta_{k,500,\rm{ul}}$  &  $f_{\eta_{k,500,\rm{ul}}}$  \\
 \hline
   11     &     1     &     0     &    NULL      &     2     &    NULL      &     2     &    87.16     &     1     &    -6.11     &     0     &    NULL      &     1    \\
   33     &     1     &     1     &    -5.41     &     0     &    NULL      &     1     &    98.78     &     1     &    -5.41     &     0     &    NULL      &     1    \\
   56     &     1     &     5     &    -4.12     &     0     &     0.02     &     0     &    94.31     &     5     &    -4.12     &     0     &     0.02     &     0    \\
   60     &     1     &     5     &    -4.94     &     0     &     0.32     &     0     &    98.16     &     5     &    -4.94     &     0     &     0.32     &     0    \\
  187     &     1     &     0     &    NULL      &     2     &    NULL      &     2     &    87.72     &     0     &    NULL      &     2     &    NULL      &     2    \\
  198     &     0     &     2     &    -2.92     &     0     &     0.15     &     0     &    93.94     &     2     &    -2.92     &     0     &     0.15     &     0    \\
  199     &     0     &     4     &    -3.78     &     0     &    -0.12     &     0     &    86.21     &     4     &    -3.78     &     0     &    -0.12     &     0    \\
\ldots & \ldots & \ldots & \ldots & \ldots & \ldots & \ldots & \ldots  & \ldots & \ldots & \ldots & \ldots & \ldots \\
\hline   
\end{tabular}
\tablefoot{The full table is available in electronic form at http://amiga.iaa.es and in CDS. The columns correspond to (1) the galaxy identification according to the CIG; (2) the result of the CIG isolation criteria for neighbours within 1\,Mpc and $|\Delta\,\varv|~\leq~500$\,km\,s$^{-1}$: ''1'' if the galaxy passes, ''0'' if it fails; (3) $k_{500}$, number of neighbours within 1\,Mpc and $|\Delta\,\varv|~\leq~500$\,km\,s$^{-1}$; (4) $Q_{500}$, tidal strength estimation using neighbours within 1\,Mpc and $|\Delta\,\varv|~\leq~500$\,km\,s$^{-1}$; (5) $f_{Q_{500}}$, flag in $Q_{500}$:''0'' if $k_{500} \geq 1$, ''2'' if $k_{500} = 0$; (6) $\eta_{k, 500}$, local number density using neighbours within 1\,Mpc and $|\Delta\,\varv|~\leq~500$\,km\,s$^{-1}$; (7) $f_{\eta_{k,500}}$, flag in $\eta_{k, 500}$: ''0'' if $k_{500} \geq 2$, ''1'' if $k_{500} = 1$, ''2'' if $k_{500} = 0$; (8) $z_{\rm{comp}}$, redshift completeness in the field; (9) $k_{500,\rm{ul}}$, number of neighbours within 1\,Mpc and $|\Delta\,\varv|~\
leq~500$\,km\,s$^{-1}$ using upper limits; (10) $Q_{500,\rm{ul}}$, tidal strength upper limit estimation using neighbours within 1\,Mpc and $|\Delta\,\varv|~\leq~500$\,km\,s$^{-1}$; (11) $f_{Q_{500,\rm{ul}}}$, flag in $Q_{500,\rm{ul}}$:''0'' if $k_{500,\rm{ul}} \geq 1$, ''2'' if $k_{500,\rm{ul}} = 0$; (12) $\eta_{k,500,\rm{ul}}$, local number density upper limit using neighbours within 1\,Mpc and $|\Delta\,\varv|~\leq~500$\,km\,s$^{-1}$; (13) $f_{\eta_{k,500,\rm{ul}}}$, flag in $\eta_{k,500,\rm{ul}}$: ''0'' if $k_{500,\rm{ul}} \geq 2$, ''1'' if $k_{500,\rm{ul}} = 1$, ''2'' if $k_{500,\rm{ul}} = 0$.}
\end{table*}

\subsubsection{Spectroscopic isolation parameters} \label{Sec:resultspecparam}

The available redshift data allowed us to calculate the two isolation parameters, the local number density and tidal strength, using physical size and physical projected distance. 

We estimated the isolation parameters $\eta_{k,500}$ and $Q_{500}$, from Eqs.~\ref{Eq:etak} and \ref{Eq:Q2012tot} respectively, taking into account all the neighbour galaxies within 1\,Mpc and $|\Delta\,\varv| \leq 500$\,km\,s$^{-1}$ with respect to the central CIG galaxy. To compare this with the photometric estimate, we also calculated $Q_{\rm{Kar, s}}$ and $\eta_{k,\rm{s}}$ only including similar-size galaxies with spectroscopy within a factor 4 in apparent diameter in 1\,Mpc.
 
We calculated the isolation parameters for the 411 CIG fields considered in the spectroscopic revision, with more than 80\% completeness in redshift. If they were incomplete, we estimated upper limits using photometric redshifts also available in the SDSS.

The values of the isolation parameters are listed in Table~\ref{Tab:resultspec}.

\section{Discussion} \label{Sec:diss}
\subsection{Photometric study} \label{Sec:dissphoto}
\subsubsection{Photometric revision of the CIG isolation criteria} \label{Sec:dissphotokara}

\begin{figure*}
\centering
\includegraphics[width=\textwidth]{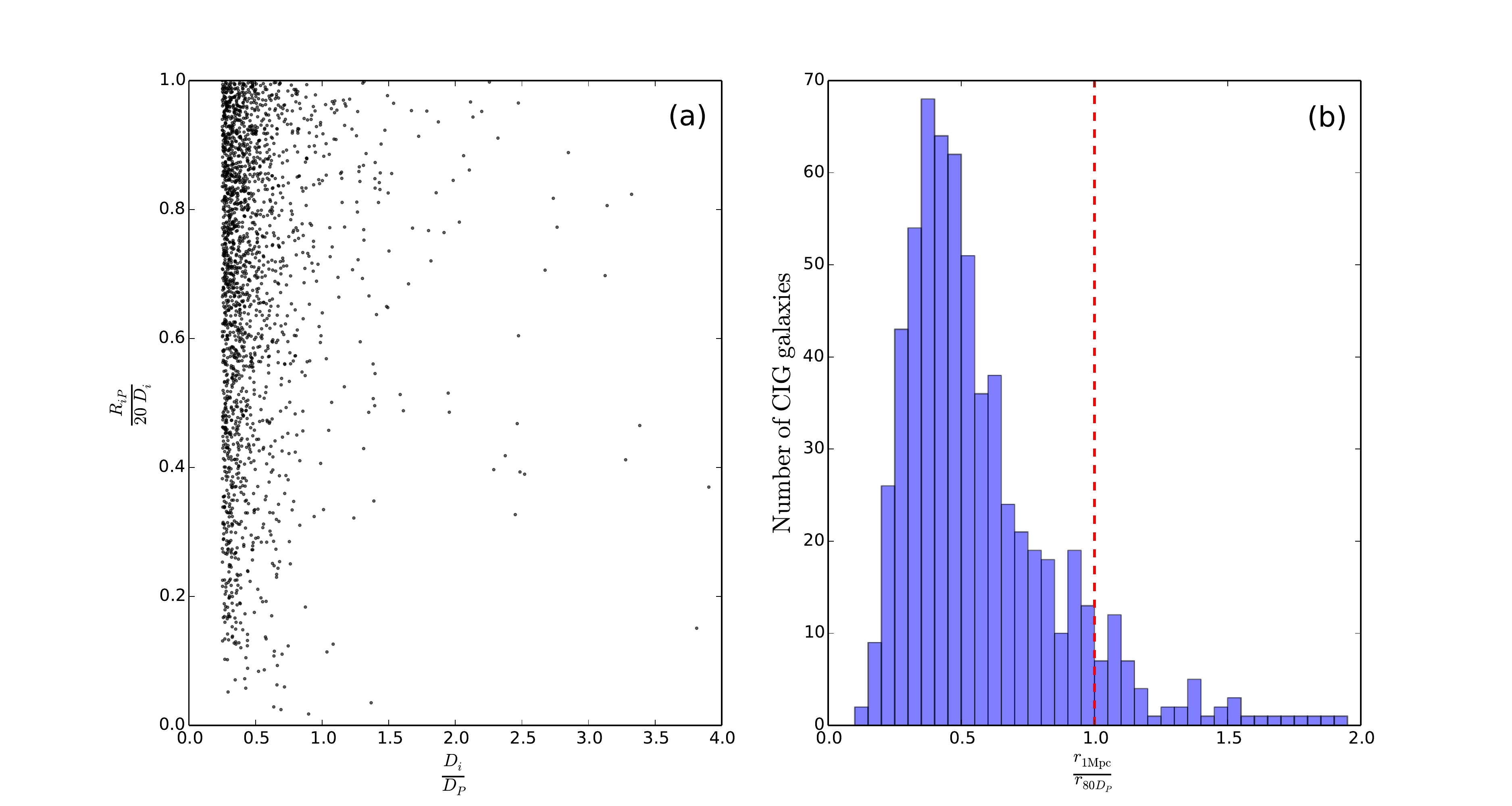}
\caption[Visualisation of photometric results]{{\it (a):} Characterisation of the 2,020 potential companions violating Eqs.~\ref{Eq:kara2} and \ref{Eq:kara1} within 1\,Mpc. {\it (b):} Distribution of the ratio between a projected angular radius at 1\,Mpc (in arcmin) across the original field radius used by \citet{1973AISAO...8....3K} ($\frac{r_{1\rm{Mpc}}}{r_{80D_{P}}}$). The vertical line corresponds to the reference value at $r_{1\rm{Mpc}}=r_{80D_{P}}$, i.e. the search area of 1\,Mpc radius is equal to the original search area used in the construction of the CIG.}
\label{Fig:photokara}
\end{figure*}

As mentioned in Sect.~\ref{Sec:resultphoto}, we performed a photometric revision of the CIG isolation criteria around each CIG galaxy within a projected radius of 1\,Mpc, finding that 2,020 neighbour galaxies are potential companions that violate Eqs.~\ref{Eq:kara2} and \ref{Eq:kara1}. The left panel in Fig.~\ref{Fig:photokara} shows that these potential neighbours tend to be smaller than their corresponding CIG galaxy and tend to concentrate at larger distances to the central galaxy. This means that small neighbours ($\frac{D_{i}}{D_{P}}<0.25$) can be located at closer distances to the CIG since their effect on the evolution of the central galaxy is almost negligible by the CIG isolation criteria. In contrast, larger neighbours could be located at gradually larger distances. That is the reason why we need to estimate the isolation degree, to quantify the effect of the missing neighbours on the evolution of the central CIG galaxy. The 2,020 potential companions are distributed around 550 CIG fields, of which 
55 cover the original search area used by \citet{1973AISAO...8....3K}. The right panel in Fig.~\ref{Fig:photokara} clearly shows that about 90\% of the CIG fields do not cover the original search area ($\frac{r_{1\rm{Mpc}}}{r_{80D_{P}}}\ <\ 1$).

\citet{2007A&A...470..505V} estimated that about 1/3 of the AMIGA sample (284 out of 950) fails the CIG isolation criteria within a minimum physical distance of 0.5\,Mpc. Although we were unable to search for companions within the original area used by \citet{1973AISAO...8....3K}, we can assess that about 1/6 of the sample fails the CIG isolation criteria within a fixed area of field radius of 1\,Mpc.
   
\begin{figure*}
\begin{center}
\includegraphics[width=\textwidth]{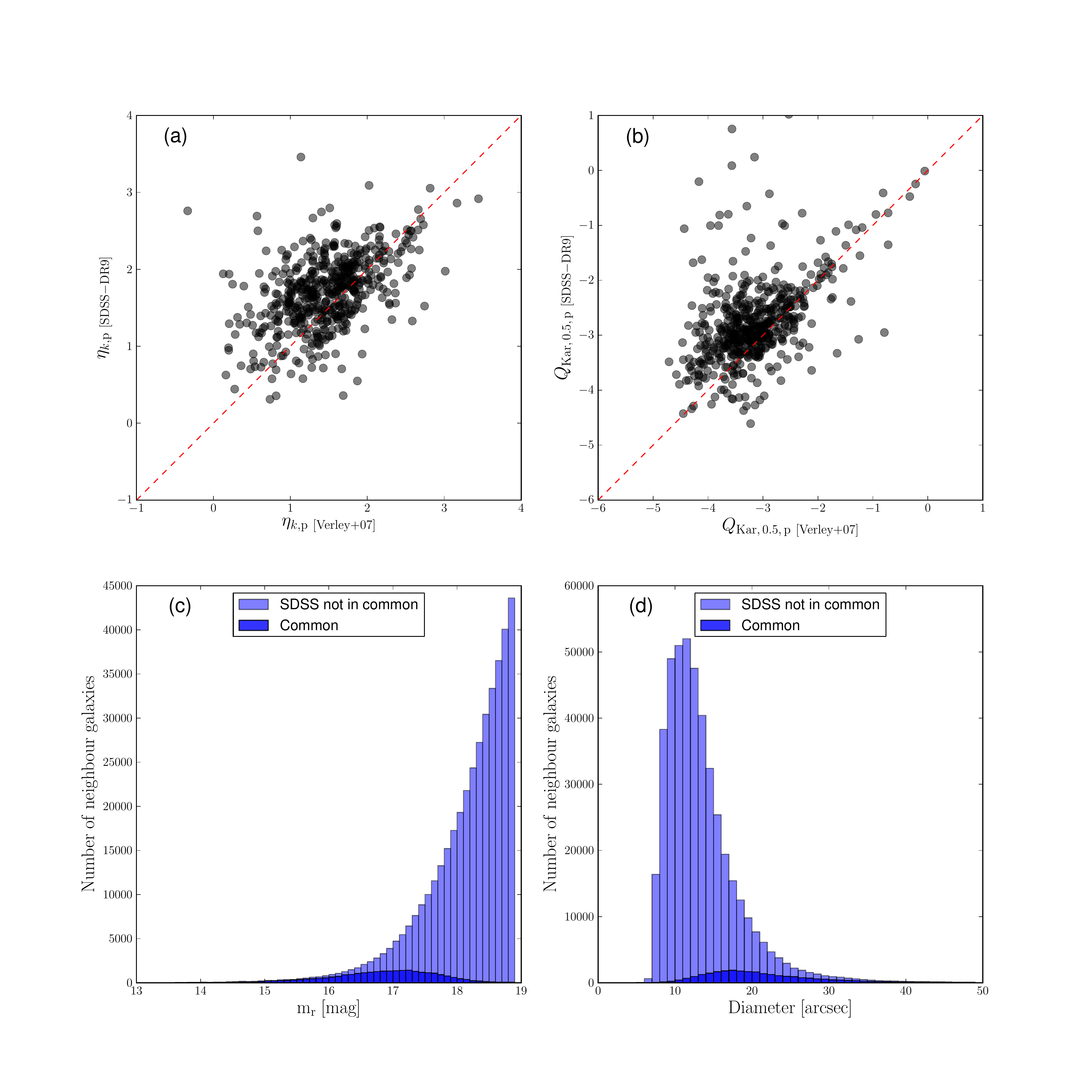} 
\end{center}
\caption[Isolation parameters in comparison to \citet{2007A&A...472..121V}]{Isolation parameter differences between this study and \citet{2007A&A...472..121V}. {\it (a):} Difference in $\eta_{k,\rm{p}}$ isolation parameter for neighbour galaxies within a factor 4 in size. {\it (b):} Difference in $Q_{\rm{0.5,Kar,p}}$ isolation parameter for neighbour galaxies within a factor 4 in size. {\it (c):} Apparent magnitude distribution for neighbour galaxies in common with \citet{2007A&A...470..505V} and galaxies in this study. {\it (d):} Apparent diameter distribution for neighbour galaxies in common with \citet{2007A&A...470..505V} and SDSS galaxies in this study.} \label{Fig:compverley} 
\end{figure*}

The sample of neighbour galaxies inspected originally by \citet{1973AISAO...8....3K} in the construction of the CIG is not available. Nevertheless, we can compare our results with the catalogue of neighbours compiled by \citet{2007A&A...470..505V}, who revised the CIG on the same original material (Palomar Observatory Sky Survey, POSS). Compared with \citet{2007A&A...470..505V}, we found a much larger number of neighbours around each CIG galaxy. Indeed, \citet{2007A&A...470..505V} extracted neighbour galaxies brighter than $B = 17.5$\,mag. In the c and d panels of Fig.~\ref{Fig:compverley}, we show that the SDSS identification of neighbours goes deeper than the POSS and also detected smaller neighbours. We also found that the 2,020 potential companions according to the CIG isolation criteria within 1\,Mpc are mostly faint, with $\Delta m_{r}\geq3$\,mag, which suggests that they are nearby and low-luminosity galaxies missed by the CIG isolation criteria. The POSS search for companions misses the faintest and 
smaller galaxies, with respect to the magnitudes of the primary CIG galaxies. In fact, the mean magnitude difference $\Delta m_{r}$ and size ratio $\frac{D_{P}}{D_{i}}$ between neighbour and the central CIG galaxy is 1.58\,dex fainter and $0.18$\,dex lower in the SDSS than in the POSS. The presence of faint galaxies does not violate the CIG isolation criteria because these systems are smaller than $1/4\times D_{P}$. The SDSS also has a redshift incompleteness at lower magnitudes $m_{r,\rm{Petrosian}}~<~14.5$\,mag \citep{2002AJ....124.1810S}. After a visual inspection of the neighbours in common with \citet{2007A&A...470..505V} that were missed by the SDSS search, we estimate that at apparent magnitudes $m_{r} < 15$\,mag, we missed, approximately one galaxy per field. These missing neighbours are usually projected close to saturated stars, which were not considered in our selection of neighbours in the SDSS.
   
Other studies about isolated galaxies claim that an equivalent CIG isolation criteria could be obtained by selecting neighbours within a magnitude range. \citet{2003ApJ...598..260P} modified Eq.~\ref{Eq:kara2} by selecting a magnitude difference of $\Delta m=2$, \citet{2010AJ....139.2525H} and \citet{2011ApJ...732...92T} considered neighbours within a magnitude difference of $\Delta m=2.5$, which translates into a factor 10 in brightness, and \citet{2005AJ....129.2062A}, the most restrictive, selected neighbours within a magnitude difference of $\Delta m=3$, which is about a factor of 16 in brightness. If we replace in Eq.~\ref{Eq:kara2} the approximate factor 4 in size with a factor 3 in magnitude, we find that 231 CIG galaxies appear to be isolated instead of 86 galaxies (see Sect.~\ref{Sec:resultphotokara}). The CIG isolation criteria are thus more restrictive and consider very faint galaxies as possible minor companions. Although the two definitions in the search for neighbours are not fully equivalent 
\citep[see][]{2007A&A...470..505V}, we found that 65\% of neighbour galaxies that violate Eq.~\ref{Eq:kara2} within 1\,Mpc have $\Delta m_{r}\geq3$, hence are low-mass objects; this means that we are able to observe faint associated satellite galaxies. This result justifies the need to quantify the isolation degree using the isolation parameters. The quantification of the tidal strength takes into account the size of the neighbour, and the effect of a satellite can be different from the effect of a similar-size neighbour galaxy.
   
\subsubsection{Photometric isolation parameters} \label{Sec:dissphotoparam}

The isolation parameters, local number density ($\eta_{k,\rm{p}}$), and tidal strength that affect the CIG galaxy ($Q_{\rm{Kar, p}}$) were estimated using photometric data (see Sect.~\ref{Sec:resultphotoparam}).
   
These two parameters are complementary in quantifying the isolation degree and give consistent results, as shown in Fig.~\ref{Fig:photoparam}a. When a galaxy presents low values for both the local number density and the tidal strength estimate, the galaxy is well isolated from any sort of external influence. In contrast, when the two values are high, the evolution of the galaxy can be perturbed by the environment, and this galaxy is not suitable to represent the normal features of isolated galaxies. Galaxies in denser environments, such as isolated pairs or triplets (see Fig.~\ref{Fig:photoparam}b), typically present relatively low values for the local number density, but high tidal strength. Studies that only use a density estimator can misclassify interacting galaxies as isolated because they do not take into account the mass of the neighbour galaxy, therefore another complementary parameter (the tidal strength) is needed. On the other hand, if the local number density is high and the tidal strength low, 
the environment of the galaxy is composed of nearby small neighbours. 

The CIG isolation criteria are also represented in Fig.~\ref{Fig:photoparam}a. The most isolated CIG galaxies, without a neighbour and depicted by black pluses, show the lowest values for both parameters. With a growing number of neighbours, CIG galaxies move to the upper right in the diagram. However, a galaxy apparently not isolated might appear in the lower left part if its $k^{\rm{th}}$ nearest neighbour is far away from the CIG and if it does not have many similar-size neighbours. CIG galaxies that fail Eqs.~\ref{Eq:kara2} and \ref{Eq:kara1} within 1\,Mpc and have a higher number of potential neighbours represent a population more that strongly interacts with their environment.

According to numerical simulations, the evolution of a galaxy may be affected by external influence when the corresponding tidal force amounts to 1\% of the internal binding force \citep{1984PhR...114..321A,1992AJ....103.1089B}, that is, $\frac{F_{\rm{tidal}}}{F_{\rm{bind}}} = 0.01$, which corresponds to a tidal strength of $Q = -2$. For the local number density, this approximately translates into a value of $\eta_{k,\rm{p}}=2.7$ (see Fig.~\ref{Fig:photoparam}a). Note that the limit value for the local number density differs from previous AMIGA works \citep[$\eta_{k}=2.4$ in][]{2007A&A...472..121V}. This theoretical value allows us to separate the interactions that might affect the evolution of the primary galaxy. Figure~\ref{Fig:photoparam}a shows that the whole subsample of 86 CIG galaxies isolated according to the CIG isolation criteria within 1\,Mpc (represented by black pluses) satisfies the threshold $Q_{\rm{Kar, p}} < -2$. 
   
Of the 550 CIG galaxies that violate the CIG isolation criteria within 1\,Mpc, 433 CIG galaxies have $Q_{\rm{Kar, p}} < -2$, and 340 CIG galaxies also have a relatively low number density environment ($\eta_{k,\rm{p}}<2.7$), therefore they can be considered to be mildly affected by their environment. Hence, from the photometric study, 426 CIG galaxies are suitable to represent a reference sample of isolated galaxies (67\% of the sample of CIG galaxies found in the photometric catalogue of the SDSS), since their evolution is dominated by internal processes.

Figure~\ref{Fig:photoparam}\,b shows the comparison of the local number density and tidal strength estimate for the CIG and for galaxies in denser environments: KPG, KTG, HCG, and ACO. Both estimates of the parameters increase from isolated galaxies to denser environments. 

These results show that the isolation parameters, even for photometric studies suffering projection effects, are sensitive enough to distinguish between environments dominated by different numbers of similar size galaxies. Quantitatively, it is important to note that the mean values of the tidal strength for denser environment are, at least, one dex higher than $Q=-2$, which means that their evolution is clearly affected by their environment.

To compare the quantification of the isolation in this study with that of \citet{2007A&A...472..121V}, we performed another calculation of the isolation parameters, restricting our fields to 0.5\,Mpc (which is the minimum physical radius used in the previous AMIGA work). When comparing tidal strengths calculated using Eqs.~\ref{Eq:Q2007} and ~\ref{Eq:Q2012}, we found a good correlation (with a systematic shift of nearly 0.5\,dex) with a large scatter. This scatter is directly related to the differences found between neighbours of the databases used, explained in Sect.~\ref{Sec:dissphotokara}. \citet{2007A&A...472..121V} provided a final catalogue of 791 isolated galaxies, based on an estimate of the best limits for selecting the sample ($Q~<~-2$ and $\eta_{k}~<~2.4$), of which 620 galaxies are in common with the present study. Of these, 486 also fulfil the new selection criteria defined in this study, hence, despite the poor correlation between the isolation parameters, only 22\% of these galaxies fail the 
new selection criteria defined here. 
   
In general, the galaxies in the samples studied here appear to be less isolated according to the new method and data. Mean values of the isolation parameters for galaxies in the CIG, KTG, HCG, and ACO, are higher (except for the tidal strength for HCG and ACO) than in previous AMIGA works \citep[see Table 8 in][compared with Table~\ref{tab:compdenser}]{2007A&A...472..121V}, which means fewer isolated galaxies. This result is directly related to the number of neighbours found in the SDSS compared with the POSS. We consider our modification for the tidal strength as a better estimate because it is based on the less scattered mass-luminosity relation. The SDSS provides linear photometric data (CCD), higher sensitivity, and better resolution than digitised photographic plates.

\subsection{Spectroscopic study} \label{Sec:dissspec}
\subsubsection{Spectroscopic revision of the CIG isolation criteria} \label{Sec:dissspeckara}

As obtained in Sect.~\ref{Sec:resultspeckara} of 347 CIG galaxies, out of 411 fields with redshift completeness higher than 80\% fulfil the CIG isolation criteria within 1\,Mpc and $|\Delta\,\varv|~\leq~500$\,km\,s$^{-1}$ when the redshift is taken into account.
   
The first isolation criterion of the CIG proposed by \citet{1973AISAO...8....3K} to remove fore- and background galaxies (Eq.~\ref{Eq:kara2}), is not fully efficient. About 50\% of the neighbours, considered as potential companions using Eq.~\ref{Eq:kara2} within 1\,Mpc, have very high recession velocities with respect to the central CIG galaxy, so the first isolation criterion of the CIG is too restrictive and could consider galaxies as not isolated that are mildly affected by their environment (see Fig.~\ref{Hist:diffvel}). On the other hand, the first isolation criterion of the CIG which requires similar apparent diameter companions, accounts for most of the physical neighbours. But we also found that about 92\% of the neighbour galaxies with recession velocities similar to the corresponding CIG galaxy are not considered as potential companions by the CIG isolation criteria.
   
We considered a different isolation criterion using the spectroscopic data, taking into account only neighbour galaxies within 1\,Mpc and $|\Delta\,\varv| \leq 500$\,km\,s$^{-1}$ with respect to the velocity of the central galaxy, that is, without imposing any difference in size. We found that 105 CIG galaxies have no physical companions instead of 347 when considering the CIG isolation criteria within 1\,Mpc and $|\Delta\,\varv|~\leq~500$\,km\,s$^{-1}$ (see Sect.~\ref{Sec:resultspeckara}). In this case we were indeed too restrictive. According to Fig.~\ref{Hist:diffvel}, we can consider that only neighbours in the peak of the distribution are physical companions of their corresponding CIG galaxy. In this case, nearly a third of the CIG sample (126 galaxies) have no physical companions (within 1\,Mpc and $|\Delta\,\varv| \leq 250$\,km\,s$^{-1}$). This means that nearby dwarf galaxies linked to the corresponding CIG galaxy were not taken into account by the CIG isolation criteria. But also part of the similar 
redshift neighbours might be background galaxies that do not affect the central CIG galaxy. We were able to recover the brightest dwarfs in the spectroscopic study, down to the $m_{r}=17.77$ magnitude limit for SDSS spectra. A more extended study will be performed in a future work, taking into account nearby and similar redshift companions to identify physical satellites that affect the evolution of the central CIG galaxy and, by consequence, a more physical estimate of the isolation degree of the CIG.

\subsubsection{Spectroscopic isolation parameters} \label{Sec:dissspecparam}

The isolation parameters local number density ($\eta_{k,500}$) and tidal strength ($Q_{500}$) were estimated for the 411 fields considered in the spectroscopic study with a redshift completeness higher than 80\% at $m_{r}=17.7$\,mag within 1\,Mpc (see Sect.~\ref{Sec:resultspec}).

Redshift information is necessary to reject fore- and background galaxies, which reduces projection effects.

\begin{figure}
\begin{center}
\includegraphics[width=0.49\textwidth]{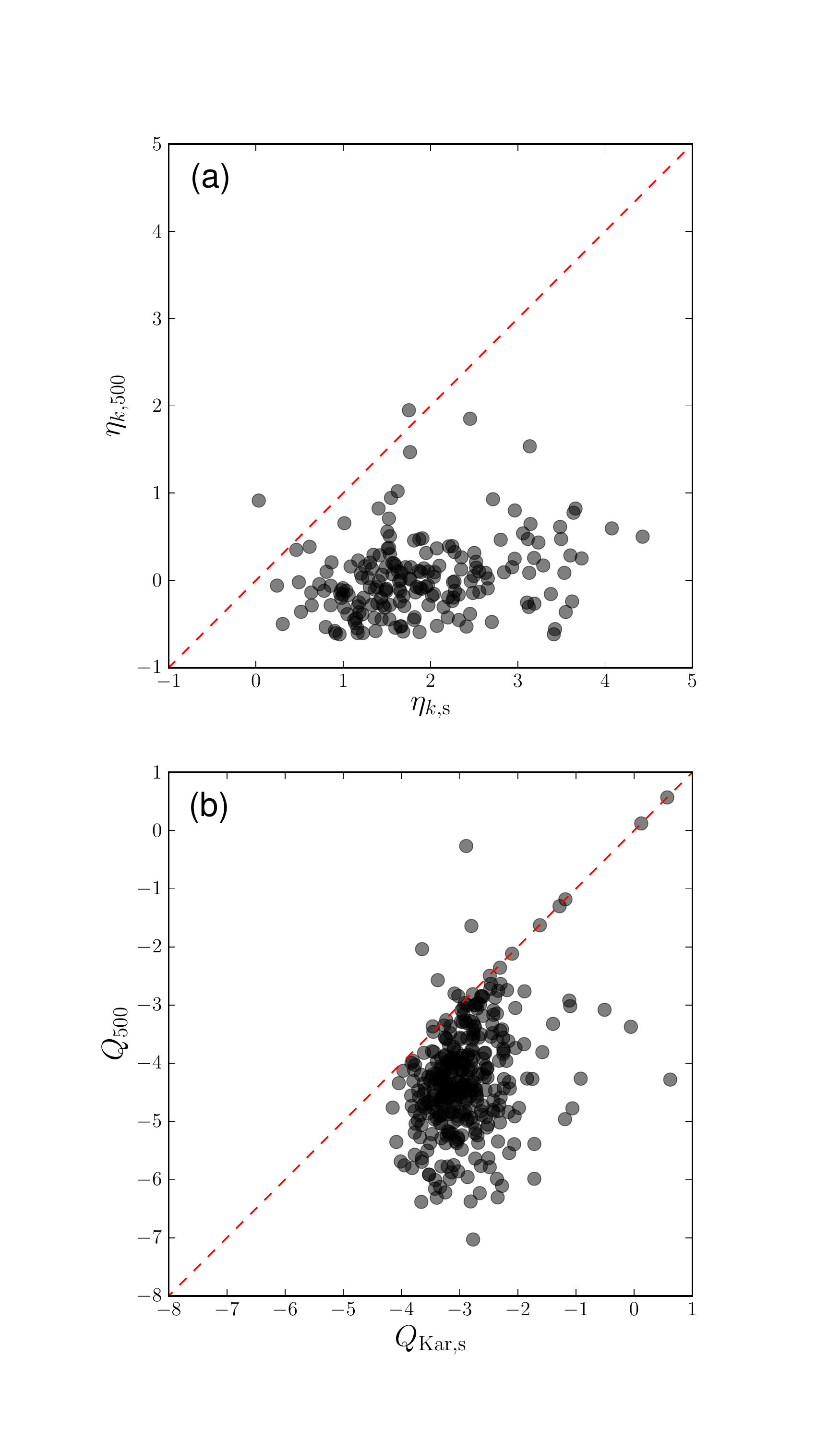} 
\end{center}
\caption[Photometric {\it vs.} spectroscopic estimates of the isolation parameters]{Photometric {\it vs.} spectroscopic estimates of the isolation parameters when using the velocity differences to reject fore- and background galaxies (vertical axis) instead of galaxies within a factor 4 in size (horizontal axis). {\it (a):} Difference in the local number density estimate. {\it (b):} Difference in the tidal strength estimate.} \label{Fig:specparam}
\end{figure}

There is no correlation between the photometric and spectroscopic estimates (see Fig.~\ref{Fig:specparam}). In general, we were unable to predict the spectroscopic parameters from the photometric estimate. Overall, the values of the isolation parameters are in general much lower in the spectroscopic estimate, showing that the projection effects lower the number of isolated candidates in the photometric study.

The upper-limit estimates of the isolation parameters were calculated considering photometric redshifts, as explained in Sect.~\ref{Sec:resultspecparam}. When the added neighbour is small and close to the CIG, the local number density changes, but the tidal strength remains almost the same. But if the neighbour is similar in size, there are marked increments in both parameters.
  
\begin{figure}
\centering
\includegraphics[width=0.49\textwidth]{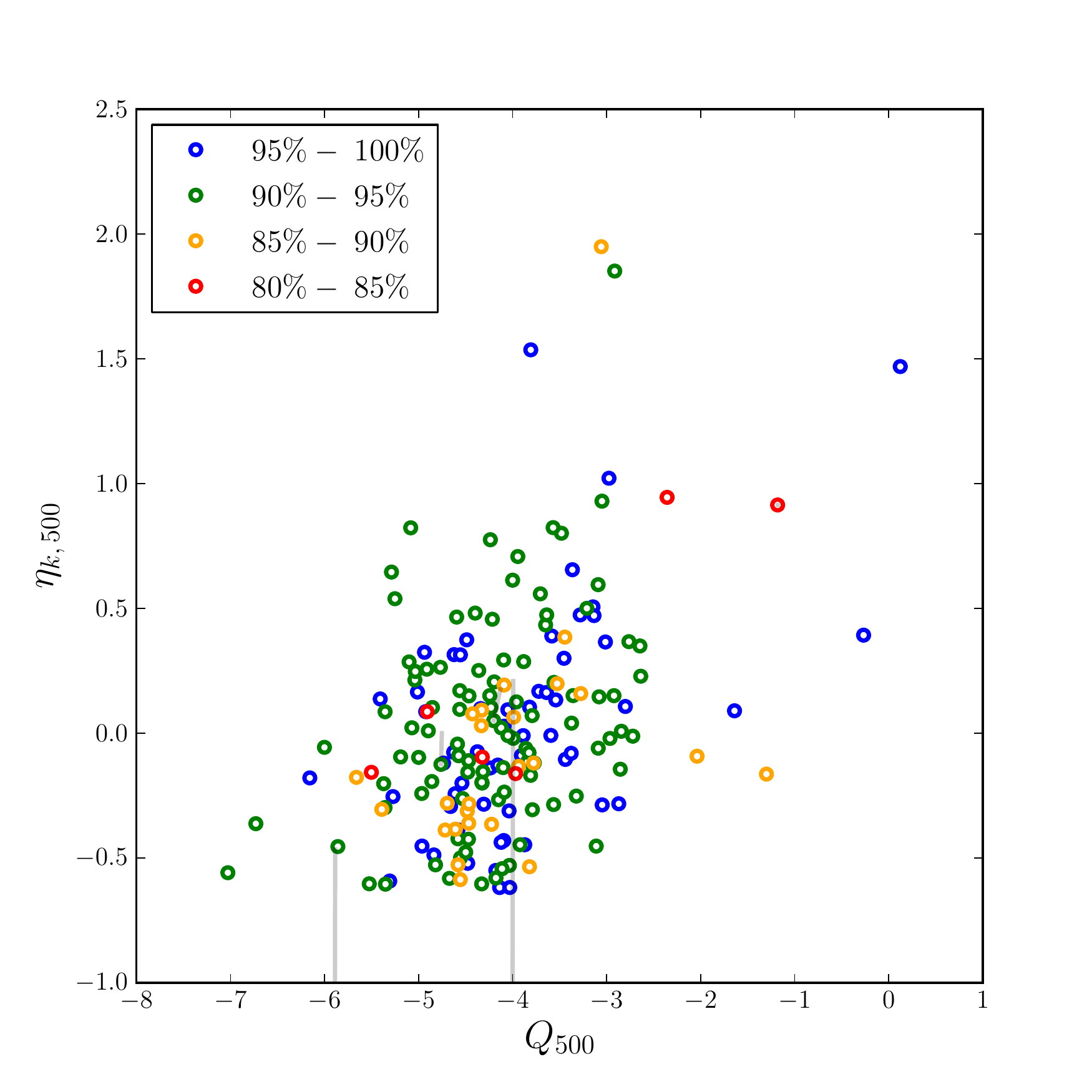}
\caption[Spectroscopic isolation parameters]{Estimate of the isolation parameters, local number density $\eta_{k,500}$ and tidal strength $Q_{500}$, for 306 CIG galaxies with at least one neighbour within 1\,Mpc and $|\Delta\,\varv|~\leq~500$\,km\,s$^{-1}$ using the spectroscopic data. Upper-limit estimates are depicted by solid grey lines. Colours, according to the legend, correspond to the redshift completeness of each CIG field (the percentage of extended neighbours, down to $m_r<17.7$\,mag lying within a  projected separation of 1\,Mpc from the CIG galaxy with a measured redshift).}
\label{Fig:specparamdzUL}
\end{figure}

The displacement due to the upper limits, represented by solid grey lines in Fig.~\ref{Fig:specparamdzUL}, is independent of the redshift completeness. Only ten CIG galaxies show changes in the parameters, the highest for CIG 492 with an increase of 0.07\,dex in the tidal strength and 0.12\,dex in the local number density. This change is due to the addition of one close ($R_{iP} \simeq 470$\,kpc) and faint ($\Delta\,m_{r}\geq-3.3$) companion with $|\Delta\,\varv| \simeq 500$\,km\,s$^{-1}$. CIG 254 and CIG 418 show an increase in the tidal strength of 0.53\,dex and 0.47\,dex, respectively. The local number density in these cases changes from being flagged as $-99$ to $\eta_{k,500} = 0.21$  and $\eta_{k,500} = -0.46$, respectively, due to the addition of a first nearest neighbour. We conclude that even if the redshift completeness of the SDSS is limited to $m_{\rm{r,Petrosian}} < 17.77$\,mag, the spectroscopic estimate of the isolation parameters is more realistic than the photometric estimate, from which the 
uncleaned objects are difficult to remove with an automated pipeline.

\subsection{Photometric versus spectroscopic studies}

We assessed the validity of some assumptions that were used during the construction of the CIG and, in light of the SDSS-DR9 spectroscopic information, considered and systematically quantified the differences between the photometric and spectroscopic studies. For the first time, we thus highlighted the quantified differences, strengths, and weaknesses of the two approaches and applied them to one common sample.

Clearly the spectroscopic information provides a better physical view of the environment of the galaxies. Nevertheless, there is still not complete full spectroscopic coverage for the neighbour galaxies of the CIG, therefore a pure photometric estimation is still needed to obtain a lower limit of the isolation parameters that is homogeneously defined and consistent for the whole CIG in the SDSS footprint. In addition, since the original materials were very different between \citet{2007A&A...470..505V}, that is, digitised photographic POSS I - II plates, and our work, it has been necessary to repeat the photometric estimation of the isolation parameters. From this we were able to perform a fair comparison between photometric and spectroscopic isolation parameters, without being biased by the discrepancies in the constructed databases (star-galaxy separation, magnitudes, completeness limits, sizes, etc).

We found that of the 411 CIG galaxies with more than 80\% redshift completeness (i.e., the percentage of extended neighbours down to $m_r<17.7$\,mag that lie within a  projected separation of 1\,Mpc from the CIG galaxy with a measured redshift), 54 CIG galaxies previously identified as isolated according to the CIG isolation criteria within 1\,Mpc actually have similar redshift companions (i.e., neighbours with $|\Delta\,\varv|~\leq~500$\,km\,s$^{-1}$ within 1\,Mpc). When considering only neighbours with $|\Delta\,\varv|~\leq~500$\,km\,s$^{-1}$ within 1\,Mpc, we find that 105 CIG galaxies of 411 show no similar redshift neighbours. There are only nine CIG galaxies isolated according to these two modifications of the CIG isolation criteria (CIG 314, 451, 473, 541, 545, 608, 613, 655, and 668).


\section{Summary and conclusions} \label{Sec:con}
   
We used the SDSS-DR9 photometric and spectroscopic databases to re-evaluate the degree of isolation of 636 galaxies in the Catalogue of Isolated Galaxies \citep[CIG; ][]{1973AISAO...8....3K}. This re-evaluation using CCD images and spectra continues and improves the work of \citet{2007A&A...472..121V,2007A&A...470..505V} which was based upon the digitised photographic plates from POSS-1 and POSS-2. We used the SDSS-DR9 to search for neighbour galaxies within a projected physical radius of 1\,Mpc, which doubles the radius used in previous AMIGA works. We first applied the CIG isolation criteria within 1\,Mpc to the SDSS photometric database. Using the SDSS spectroscopic database, we then refined the study for 411 fields, of which more than 80\% of the extended neighbours down to $m_r<17.7$\,mag lying within a projected separation of 1\,Mpc from the CIG galaxy have a measured redshift. The isolation degree was quantified using two different and complementary parameters: the local number density $\eta_{k}$ and 
the tidal strength $Q$, which affect the central CIG galaxy. 

A summary of the different samples used in the photometric and spectroscopic studies is shown in Table~\ref{tab:summarysamples}.
   
\begin{table*}
  \caption[Summary of samples used in the study]{\textbf{Summary of the samples used in the photometric and spectroscopic studies.}}
   \label{tab:summarysamples}
\centering
\begin{tabular}{cl}
\hline \hline
Number of entries & Definition of the sample \\
\hline 
1050 & CIG galaxies in the original catalogue (Karachentseva 1973) \\
799  & CIG galaxies found in the photometric catalogue of the SDSS-DR9 \\
789  & CIG galaxies of 799 after deleting of 10 galaxies with unreliable photometric data from the SDSS-DR9 \\
693  & CIG galaxies of 789 after deleting of 96 galaxies with $\varv < 1500$\,km\,s$^{-1}$ \\
636  & CIG galaxies of 693 after deleting of 57 galaxies with a field radius of 1\,Mpc not covered in the photometric\\
 & SDSS-DR9 catalogue \\
411  & CIG galaxies of 636 of which more than 80\% have extended neighbours down to $m_r<17.7$\,mag that lie within \\
 & a projected separation of 1\,Mpc from the CIG galaxy with a measured redshift \\
\hline \hline
86   & CIG galaxies of 636 without neighbours that violate Eqs.~\ref{Eq:kara2} and \ref{Eq:kara1} within 1\,Mpc \\
117  & CIG galaxies of 636 without neighbours that violate Eqs.~\ref{Eq:kara2} and \ref{Eq:kara1} within 1\,Mpc \\
433  & CIG galaxies of 636 with more than one neighbour that violate Eqs.~\ref{Eq:kara2} and \ref{Eq:kara1} within 1\,Mpc \\
550  & CIG galaxies of 636 (117+433), with neighbours that violate Eqs.~\ref{Eq:kara2} and \ref{Eq:kara1} within 1\,Mpc \\
231  & CIG galaxies of 636 without neighbours that violate Eq.~\ref{Eq:kara1} and for which, additionally, the approximate factor 4 \\
 & in size was replaced in Eq.~\ref{Eq:kara2} by a factor 3 in magnitude within 1\,Mpc \\
\hline \hline
433  & CIG galaxies of 550 with a tidal strength $Q_{\rm{Kar, p}} < -2$ \\
340  & CIG galaxies of 550 with a tidal strength $Q_{\rm{Kar, p}} < -2$ and a local number density $\eta_{k,\rm{p}}<2.7$ \\
426  & CIG galaxies of 636 (86+340) without neighbours that violate Eqs.~\ref{Eq:kara2} and \ref{Eq:kara1} within 1\,Mpc \\
 & and with tidal strength $Q_{\rm{Kar, p}} < -2$ and local number density $\eta_{k,\rm{p}}<2.7$ \\
\hline \hline
347  & CIG galaxies of 411 without neighbours that violate Eqs.~\ref{Eq:kara2} and \ref{Eq:kara1}, and with $|\Delta\,\varv|~\leq~500$\,km\,s$^{-1}$ within 1\,Mpc \\
105  & CIG galaxies of 411 without neighbours with $|\Delta\,\varv|~\leq~500$\,km\,s$^{-1}$ within 1\,Mpc \\
308  & CIG galaxies of 411 with at least one neighbour with $|\Delta\,\varv|~\leq~500$\,km\,s$^{-1}$ within 1\,Mpc \\
\hline \hline
\end{tabular}
\end{table*} 

Our conclusions are the following: 
   
\begin{enumerate}
\item Of the 636 CIG galaxies considered in the photometric study, 426 galaxies appear to be isolated in projection: 86 CIG galaxies are isolated according to the CIG isolation criteria within a projected field radius of 1\,Mpc; 340 appear to be mildly affected by their environment. 

\item The use of the SDSS database permits one to identify faint companions that were not found in previous AMIGA papers \citep{2007A&A...470..505V}. The SDSS provides linear photometry, improved sensitivity, and better spatial resolution than digitised photographic plates. Consequently, the isolation parameters of the revised AMIGA sample are improved, which reduces the sample by about 20\%.

\item On average, galaxies in the AMIGA sample show lower values in the local number density and the tidal strength parameters than galaxies in denser environments such as pairs, triplets, compact groups, and clusters. In general, galaxies in the studied samples show higher values of the isolation parameters than those reported by \citet{2007A&A...472..121V}. 

\item Of the 411 fields considered in the spectroscopic study with more than 80\% redshift completeness, 347 galaxies are isolated according to the CIG isolation criteria within a radius of 1\,Mpc and $|\Delta\,\varv|~\leq~500$\,km\,s$^{-1}$ with respect to the central CIG galaxy.

\item The upper-limit estimates of the isolation parameters were calculated considering photometric redshifts: 103 CIG galaxies have no neighbours within 1\,Mpc  within the specified apparent diameter range and $|\Delta\,\varv|~\leq~500$\,km\,s$^{-1}$.

\item The spectroscopic local number density and the tidal strength were calculated for 308 CIG galaxies with at least one neighbour within 1\,Mpc and $|\Delta\,\varv|~\leq~500$\,km\,s$^{-1}$. This estimate improves the quantification of the isolation degree with respect to the photometric study, which is only a rough first approximation. 

\item The availability of the spectroscopic data allowed us to check the validity of the CIG isolation criteria within a field radius of 1\,Mpc, which is not fully efficient. About 50\% of the neighbours considered as potential companions in the photometric study are in fact background objects. On the other hand, we also found that about 92\% of neighbour galaxies that show recession velocities similar to the corresponding CIG galaxy are not considered by the CIG isolation criteria as potential companions. These neighbours are most likely dwarf systems, with $D_{i}\ <\ 0.25\ D_{P}$, which may have a considerable influence on the evolution of the central CIG galaxy.
\end{enumerate}


\begin{acknowledgements}   
The authors acknowledge the referee for his/her very detailed and useful report, which helped to clarify and improve the quality of this work.

This work has been supported by Grant AYA2011-30491-C02-01, co-financed by MICINN and FEDER funds, and the Junta de Andalucía (Spain) grants P08-FQM-4205 and TIC-114, as well as under the EU 7th Framework Programme in the area of Digital Libraries and Digital Preservation. (ICT-2009.4.1) Project reference: 270192.

This work was partially supported by a Junta de Andalucía Grant FQM108 and a Spanish MEC Grant AYA-2007-67625-C02-02.

Funding for SDSS-III has been provided by the Alfred P. Sloan Foundation, the Participating Institutions, the National Science Foundation, and the U.S. Department of Energy Office of Science. The SDSS-III web site is http://www.sdss3.org/. SDSS-III is managed by the Astrophysical Research Consortium for the Participating Institutions of the SDSS-III Collaboration including the University of Arizona, the Brazilian Participation Group, Brookhaven National Laboratory, University of Cambridge, University of Florida, the French Participation Group, the German Participation Group, the Instituto de Astrofisica de Canarias, the Michigan State/Notre Dame/JINA Participation Group, Johns Hopkins University, Lawrence Berkeley National Laboratory, Max Planck Institute for Astrophysics, New Mexico State University, New York University, Ohio State University, Pennsylvania State University, University of Portsmouth, Princeton University, the Spanish Participation Group, University of Tokyo, University of Utah, Vanderbilt 
University, University of Virginia, University of Washington, and Yale University.

This research has made use of data obtained using, or software provided by, the UK's AstroGrid Virtual Observatory Project, which is funded by the Science and Technology Facilities Council and through the EU's Framework 6 programme. We also acknowledge the use of STILTS and TOPCAT tools \citet{2005ASPC..347...29T}.

This research made use of python ({\tt http://www.python.org}), of Matplotlib \citep{Hunter:2007}, a suite of open-source python modules that provides a framework for creating scientific plots.

This research has made use of the NASA/IPAC Extragalactic Database (NED), which is operated by the Jet Propulsion Laboratory, California Institute of Technology, under contract with the National Aeronautics and Space Administration. 

We acknowledge the usage of the HyperLeda database ({\tt http://leda.univ-lyon1.fr}) \citep{2003A&A...412...45P}.
\end{acknowledgements}

\bibliography{astroph}

\end{document}